\pdfoutput=1
\documentclass{aa2}

\usepackage{graphicx}
\usepackage{txfonts}

\usepackage[hidelinks]{hyperref}
\renewcommand{\d}{\mathrm{d}}
\newcommand{\vtheta}{\vec{\theta}}
\usepackage{tikz-cd}
\usepackage[normalem]{ulem}

\usepackage{mathtools}
\newcommand{\set}[1]{\ensuremath{ \left\lbrace #1 \right\rbrace}}
\newcommand{\mbR}{\ensuremath{\mathbb{R}}}
\newcommand{\mbT}{\ensuremath{\mathbb{T}}}
\newcommand{\mbX}{\ensuremath{\mathbb{X}}}

\newcommand{\betti}{\ensuremath{\beta}}

\newcommand{\persbetti}{\ensuremath{\beta}}
\newcommand{\relpersbetti}{\ensuremath{\beta^{M}}}
\newcommand{\dgm}{\ensuremath{\operatorname{Dgm}}}

\newcommand{\regbettiKV}{\ensuremath{b}}
\newcommand{\persbettiKV}{\ensuremath{b^{\operatorname{per}}}}

\begin{document}

\title{
Persistent homology in cosmic shear: constraining parameters with topological data analysis
}
\author{Sven Heydenreich
      \inst{1}
      \and
      Benjamin Br\"uck
      \inst{2}
      \and
      Joachim Harnois-D\'eraps
      \inst{3,4}
      }

\institute{
      Argelander-Institut f\"ur Astronomie, Auf dem H\"ugel 71, 53121 Bonn, Germany 
      \and 
      University of Copenhagen, Department of Mathematical Sciences, Universitetsparken 5, DK-2100 Copenhagen, Denmark
      \and
      Astrophysics Research Institute, Liverpool John Moores University, 146 Brownlow Hill, Liverpool L3 5RF
      \and
      Institute for Astronomy, University of Edinburgh, Royal Observatory, Blackford Hill, Edinburgh, EH9 3HJ, UK
      \\ \email{sven@astro.uni-bonn.de}
      }

\abstract
{
In recent years, cosmic shear has emerged as a powerful tool to study the statistical distribution of matter in our Universe. Apart from the standard two-point correlation functions, several alternative methods like peak count statistics offer competitive results. Here we show that persistent homology, a tool from topological data analysis, can extract more cosmological information than previous methods from the same dataset. For this, we use persistent Betti numbers to efficiently summarise the full topological structure of weak lensing aperture mass maps. This method can be seen as an extension of the peak count statistics, in which we additionally capture information about the environment surrounding the maxima. We first demonstrate the performance in a mock analysis of the KiDS+VIKING-450 data: we extract the Betti functions from a suite of $w$CDM $N$-body simulations and use these to train a Gaussian process emulator that provides rapid model predictions; we next run a Markov-Chain Monte Carlo analysis on independent mock data to infer the cosmological  parameters and their uncertainty. When comparing our results, we recover the input cosmology and achieve a constraining power on $S_8 \equiv \sigma_8\!\sqrt{\Omega_\mathrm{m}/0.3}$ that is 5\% tighter than that of peak count statistics. Performing the same analysis on 100 deg$^2$ of {\it Euclid}-like simulations, we are able to improve the constraints  on $S_8$  and $\Omega_\mathrm{m}$ by  18\% and 10\%, respectively, while breaking some of the degeneracy between $S_8$ and  the dark energy equation of state. To our knowledge, the methods presented here are the most powerful topological tools to constrain cosmological parameters with lensing data.}


\maketitle

\section{Introduction}
The standard model of cosmology, $\Lambda$CDM, has been incredibly successful in explaining and predicting a large variety of cosmological observations using only six free parameters. As of now, the tightest constraints on these cosmological parameters have been placed by observations of the cosmic microwave background \citep[CMB,][]{2018arXiv180706209P}. However, the next generation of surveys like the Rubin Observatory\footnote{\url{https://www.lsst.org/}} \citep{Ivezic:2008}, \textit{Euclid}\footnote{\url{https://www.euclid-ec.org/}} \citep{Laurejis:2011} and the Nancy Grace Roman Space Telescope\footnote{\url{https://roman.gsfc.nasa.gov/}} \citep[RST,][]{Spergel:2013} promises to improve on these constraints and reduce parameter uncertainties to the sub-percent-level. This is particularly interesting in light of tensions arising between observations of the early Universe (CMB) and the late Universe at a redshift of $z\lesssim 2$. These tensions are most notable in the Hubble constant $H_0$ \citep{Riess:2019} and the parameter $S_8 = \sigma_8\sqrt{\Omega_\mathrm{m}/0.3}$ \citep{Joudaki:2020,2020arXiv200211124D}, where $\Omega_\mathrm{m}$ is the matter density parameter and $\sigma_8$ characterises the normalisation of the matter power spectrum. Future analyses will show whether these tensions are a statistical coincidence, the manifestation of unknown systematics or evidence for new physics.

In the last decade, weak gravitational lensing has proven to be an excellent tool to study the distribution of matter and constrain the cosmological parameters that describe our Universe. In particular, tomographic shear two-point correlation functions \citep{Kaiser:1992}  and derived quantities like Complete Orthogonal Sets of E- and B-mode Integrals \citep[COSEBIs,][]{Schneider:2010} or band powers \citep{vanUitert:2018} have been applied with great success to cosmic shear data; see \citet{Kilbinger:2013,arXiv:1303.1808,arXiv:1601.05786} for an analysis of  CFHTLenS, \citet{Hildebrandt:2020,arXiv:2005.04207,Asgari:2020} for the Kilo Degree Survey, \citet{DES1_Troxel} for the Dark Energy Survey and \citet{HSC_Cl,arXiv:1906.06041} for the Hyper Suprime Camera Survey.  However, while two-point statistics provide an excellent tool to capture the information content of Gaussian random fields, the gravitational evolution of the matter distribution becomes non-linear in high-density regions, and additional methods are needed to extract the information residing in the non-Gaussian features that are formed therein. The demand for these new methods rises with the ever-increasing amount and quality of available data. 

In addition to three-point correlation functions \citep{2003A&A...397..809S,Fu2014}, which form a natural extension to two-point correlation functions, a variety of alternative statistics promise to improve cosmological parameter constraints \citep{2020arXiv200612506Z}, including peak statistics
\citep[][hereafter M+18]{2015MNRAS.450.2888L, 2015PhRvD..91f3507L, DES-SV-peaks, Martinet:2018}, 
density split statistics \citep{Gruen2017,Burger:2020}, convergence probability distribution functions \citep{MassiveNu2},\linebreak shear clipping \citep{2018MNRAS.480.5529G}, the scattering transform \citep{2020arXiv200608561C}, and Minkowski functionals \citep{2015PhRvD..91j3511P, MassiveNu3,2020A&A...633A..71P}.

In this article, we demonstrate how  {\it persistent homology} can be used to analyse the data provided by weak gravitational lensing surveys. This non-linear statistics combines the information residing in Minkowski functionals and in peak statistics in a natural way, and supplements these by further capturing information about the environment surrounding the topological features. In the last two decades, persistent homology has been successfully applied in a large variety of fields involving topological data analysis: among others, it has been used to analyse the spread of contagious diseases \citep{10.1371/journal.pone.0192120}, to assess self-similarity in geometry \citep{https://doi.org/10.1063/1.4737391} and to identify a new subgroup of breast cancer \citep{10.1073/pnas.1102826108}; for more examples, we refer to \citet{arXiv:1506.08903} and references therein.

In cosmology, persistent homology has already been used to study the topology of the cosmic web \citep{2011MNRAS.414..350S,2013arXiv1306.3640V},  of the interstellar magnetic fields \citep{2018MNRAS.475.1843M} and the reionisation bubble network \citep{2019MNRAS.486.1523E}. Furthermore, it  was shown to be an effective cosmic void finder \citep{2019A&C....27...34X}. Additionally, a specific summary of the information contained in persistent homology, the {\it Betti numbers}, has emerged as a powerful tool to analyse both the cosmic web \citep{2017MNRAS.465.4281P} and Gaussian random fields \citep[][hereafter, P+19]{2019A&A...627A.163P}. In the latter case, Betti numbers  have been shown to provide a higher information content than both Euler characteristics and Minkowski functionals \citep{2019MNRAS.485.4167P}. In particular, they are very effective at detecting and quantifying  non-Gaussian features in fields  such as the CMB temperature map (compare P+19). 

While several of the above-mentioned papers have shown that Betti numbers are sensitive to cosmological parameters, their results were only qualitative so far. To our knowledge, the current article is the first to quantify the power of Betti numbers for constraining cosmological parameters. 
Furthermore, our work differs from the previous ones in that we use \emph{persistent} Betti numbers, which further take into account the environment surrounding the topological features (for the definitions and further explanations, see Sect.~\ref{sec:homology}).
We show that this leads to a significant improvement in constraining power  compared to  the {\it non-persistent} Betti numbers that have so far been used in cosmology.

In our analysis, we constrain cosmological parameters with a Markov-Chain Monte Carlo (MCMC) sampler, and therefore need to efficiently compute Betti functions over a broad range of values. Since we are not aware of a way to model Betti functions for non-Gaussian fields, we instead rely on Gaussian process regression \citep[see e.g.][]{gelmanbda04}, a machine learning tool that probabilistically predicts a given statistics when only small sets of training data are available. This procedure is not restricted to Betti functions and has been used in cosmology  for a number of other statistical methods \citep{Heitmann2013, MassiveNu2, DSS_Burger}. Since persistent homology is particularly efficient in summarising and compressing the topological structure of large datasets, it is well suited for interacting with machine-learning algorithms \citep{2019arXiv191008245B}.
 
In this work, we train a Gaussian process regressor on a suite of KiDS+VIKING-450-like \citep[][hereafter KV450-like]{Wright:2019} $N$-body simulations to predict Betti functions for arbitrary cosmological parameters within a wide training range in a  $w$CDM cosmology. Calibrating our covariance matrix from a distinct  ensemble of fully independent simulations, we then perform an MCMC analysis on mock data and recover the fiducial cosmological parameters of the simulation. We further show that persistent Betti functions are able to constrain cosmological parameters better than peak statistics, whose performance is similar to the one of tomographic cosmic shear \citep[][M+18]{DES-SV-peaks} and slightly better than that of Minkowski functionals \citep{2020arXiv200612506Z}.

We finally carry out a mock analysis based on \textit{Euclid}-like simulations and find an even larger increase in constraining power in this setup. Thus, persistent Betti numbers promise to substantially improve the constraining power of weak gravitational lensing surveys, especially when they are used in combination with other probes.

This paper is organised as follows: In Sect.~\ref{sec:methods} we give a brief overview  of the $N$-body simulations used in the signal calibration and in the estimation of the covariance matrix (Sect.~\ref{sec:simulations}), of the aperture mass maps reconstruction (Sect.~\ref{sec:map}), of  the  theory underlying persistent homology statistics (Sect.~\ref{sec:homology}) and Gaussian process regression emulation (Sect.~\ref{sec:gpr}). An explanation of the peak count statistics can be found in Sect.~\ref{subsec:peaks}. We present the results of our KV450-like and \textit{Euclid}-like analyses in Sects.~\ref{sec:results} and \ref{sec:euclid}, respectively,  and discuss them in Sect.~\ref{sec:discussion}.

\section{Methods and numerical data products}
\label{sec:methods}
Throughout this work, we assume the standard weak gravitational lensing formalism, a review of which can be found in \citet{Bartelmann:2001}.
\subsection{Weak lensing simulations}
\label{sec:simulations}

So far, we cannot analytically compute the Betti functions that describe cosmic shear data due to their highly non-linear nature. Instead, we rely on numerical simulations\footnote{All simulations products used in this paper can be made available upon request; see http://slics.roe.ac.uk}, namely the Scinet LIghtCone Simulations \citep[SLICS,][]{2018MNRAS.481.1337H,2015MNRAS.450.2857H} and the cosmo-SLICS \citep[][hereafter H+19]{2019A&A...631A.160H} for their evaluation.

 First, we extract the cosmology dependence with the cosmo-SLICS, a suite of cosmological $N$-body simulations in which the matter density $\Omega_\mathrm{m}$, the parameter $S_8$, the Hubble constant $h$ and the parameter for the dark energy equation of state $w_0$ are sampled at 26 points in a Latin hyper-cube (see Tab.~\ref{tab:cosm_param} for the exact list).  At each cosmology, a pair of  $N$-body simulations were evolved to redshift $z=0$ with $1536^3$ particles in a box of $505\,h^{-1}\,\mathrm{Mpc}$, and subsequently ray-traced multiple times to yield 50 pseudo-independent light-cones that cover each $10\times 10\,\mathrm{deg}^2$. The initial conditions were chosen such as to suppress  most of the sample variance when averaging a statistic over the pair. (For more details on the cosmo-SLICS, we refer the reader to H+19.) 

Second, we use the SLICS to estimate the covariance of the Betti functions. These consist of a set of 126 fully independent $N$-body simulations\footnote{The full SLICS ensemble contains 820 realisations, however we only use a sub-sample of 126 in this work.} conducted  in a flat $\Lambda$CDM-Universe with $\Omega_\mathrm{m}=0.2905$, $\Omega_\mathrm{b}=0.0473$, $h=0.6898$, $\sigma_8=0.826$ and $n_\mathrm{s}=0.969$.  These were also ray-traced into $10\times 10\,\mathrm{deg}^2$ light-cones, in a format otherwise identical to the cosmo-SLICS.

We take the KV450 data set \citep{Wright:2019} as an example of a current Stage-III weak lensing survey, and create mock data sets with similar properties.
Due to the box size of our simulations, the full KV450 survey footprint cannot be fitted onto a single light-cone. Instead we split the survey into 17 tiles following the setup presented in Appendix A3 of \citet{2018MNRAS.481.1337H}, and computed the (simulated) shear signal at the exact positions of the KV450 galaxies, repeating the process for 10 light-cones (out of the 50 available) for each cosmo-SLICS pair, and for the 126 SLICS realisations. We further use the observed ellipticities to simulate the shape noise, and the galaxy redshifts were randomly selected such that the cumulative redshift distribution follows the  fiducial ``direct calibration'' method (DIR) described in \citet{Hildebrandt:2020}. This way, any effect that  the galaxy density, shape noise and survey footprint may have on the Betti functions is infused in the simulations as well.

In Sect.~\ref{sec:euclid} we further use a separate set of simulations in which the redshift distribution and galaxy number density has been modified to match that of future weak lensing experiments. These {\it Euclid}-like catalogues are also based on the SLICS and cosmo-SLICS, and we provide more  details about them in that section. Note that in neither of the simulated surveys do we split the galaxy catalogues with redshift selection; we leave tomographic analysis for future work.

\subsection{Maps of aperture mass}
\label{sec:map}
We perform our topological analysis on data obtained from weak gravitational lensing  surveys, more precisely on maps of aperture mass \citep{Schneider:1996}, which reconstruct the lensing convergence $\kappa$ inside an aperture filter and are hence directly related to the projected mass density contrast. 

There are several alternative ways to reconstruct convergence maps \citep[see, e.g.][and references therein]{Kaiser:1993,Seitz:2001,Jeffrey:2018}, however, these suffer from the so-called mass-sheet degeneracy: even under ideal circumstances,  $\kappa$ can only be determined up to a constant. The introduction of a uniform mass sheet $\kappa_0$ can change the extracted signal-to-noise ratio in a way that does not reflect any physical meaning,   hence we choose to perform our analysis on aperture mass maps, which are invariant under this effect.

\subsubsection{Theoretical background}
The aperture mass  is obtained from a weak lensing catalogue as
\begin{equation}
  M_\mathrm{ap}(\vtheta) = \int \d^2\theta' Q(|\vec{\theta'}|)\gamma_\mathrm{t}(\vtheta';\vtheta)\, ,
  \label{eq:Map_definition}
\end{equation}
 where the filter function $Q$ is computed via
\begin{equation}
  Q(\theta) = \frac{2}{\theta^2}\int_0^\theta\d\theta'\,\theta'U(\theta')-U(\theta)\, ,
\end{equation}
and $U$ is a compensated filter with $\int\,\d\theta\, \theta U(\theta) = 0$  \citep{Schneider:1996}.
The tangential shear at position $\vtheta'$ with respect to $\vtheta$, $\gamma_\mathrm{t}(\vtheta';\vtheta)$, is calculated as 
\begin{equation}
  \gamma_\mathrm{t}(\vtheta';\vtheta) = -\Re\left[\gamma(\vtheta')\,\frac{(\vtheta'^* - \vtheta^*)^2}{|\vtheta' - \vtheta|^2}\right]\, ,
  \label{eq:gammat_definition}
\end{equation}
where $\gamma(\vtheta')$ is the (complex) shear at position $\vtheta'$. We note that both the shear $\gamma$ and the angular position $\vtheta$ are interpreted as complex quantities  in the above expression. In reality, we do not measure a shear field  directly, but rather ellipticities $\epsilon$ of an ensemble of galaxies  $n_\mathrm{gal}$, which are related to the shear and a measurement noise term $\epsilon_{\rm n}$ as
\begin{equation}
\epsilon = \frac{\epsilon_{\rm n} + g}{ 1 + \epsilon_{\rm n}g^*} \sim  \epsilon_{\rm n} + \gamma\, , 
 \label{eq:ellipticity_definition}
\end{equation}
where $g=\gamma/(1-\kappa)$ is the reduced shear and the last approximation holds in the weak shear limit. This transforms the integral of Eq.~\eqref{eq:Map_definition} into a sum 
\begin{equation}
  M_\mathrm{ap}(\vtheta) = \frac{1}{n_\mathrm{gal}}\sum_i Q(|\vtheta_i-\vtheta|)\epsilon_\mathrm{t}(\vtheta_i;\vtheta)\, ,
\end{equation}
where $\epsilon_\mathrm{t}(\vtheta_i;\vtheta)$ is the tangential ellipticity defined in analogy to Eq.~\eqref{eq:gammat_definition}.

Following M+18, we compute the noise in the aperture mass as
\begin{equation}
  \sigma\left(M_\mathrm{ap}(\vtheta)\right) = \frac{1}{\sqrt{2}n_\mathrm{gal}}\sqrt{\sum_i |\epsilon(\vtheta_i)|^2Q^2(|\vtheta_i - \vtheta|)}\, ,
\end{equation}
and calculate the signal-to-noise ratio at a position $\vtheta$ as
\begin{equation}
  \frac{S}{N}(\vtheta) = \frac{\sqrt{2}\sum_iQ(|\vtheta_i - \vtheta|)\,\epsilon_\mathrm{t}(\vtheta_i;\vtheta)}{\sqrt{\sum_i |\epsilon(\theta_i)|^2Q^2(|\vtheta_i-\vtheta|)}} \, .
  \label{eq:S/N_definition}
\end{equation}
Both numerator and denominator of Eq.~\eqref{eq:S/N_definition} can be expressed as a convolution and can therefore be computed via a Fast Fourier-Transform (FFT), significantly decreasing the required computation time \citep{Unruh:2019}. For our analysis, we use the following filter function \citep[see][M+18]{Schirmer:2007}
\begin{align}
  Q(\theta) = & \left[1 + \exp \left(6 -150 \frac{\theta}{\theta_{\rm ap}}\right) + \exp \left(-47 +50 \frac{\theta}{\theta_{\rm ap}}\right)\right]^{-1}
             \nonumber\\
             & \quad \times \left(\frac{\theta}{x_{\rm c}\theta_{\rm ap}}\right)^{-1} \tanh \left(\frac{\theta}{x_{\rm c}\theta_{\rm ap}}\right)\, .
             \label{eq:filterfunction}
\end{align}
This function was designed to  efficiently follow the  mass profiles of dark matter haloes, which we model according to  \citet[][NFW hereafter]{Navarro:1997}; since most of the matter is located within dark matter haloes, the function is well-suited to detect peaks in the matter distribution. Here, $\theta_\mathrm{ap}$ is the aperture radius and $x_\mathrm{c}$ represents the concentration index of the NFW profile. Following M+18, we set $x_\mathrm{c}=0.15$, which is a good value for detection of galaxy clusters \citep{Hetterscheidt:2005}.

\subsubsection{Numerical implementation}
\label{subsubsec:numerical_Map}

 The signal-to-noise ratio defined in Eq.~\eqref{eq:S/N_definition} is highly sensitive to the noise properties of the galaxy survey, and it is therefore critical to reproduce exactly the galaxy number density and the intrinsic shape noise of the data, as well as the overall survey footprint (see e.g.~M+18). This motivates  our use of KV450 mosaic simulations, which reproduce all of these quantities exactly.  For each simulated galaxy in the catalogue, we randomly rotate the observed ellipticity of the  corresponding real galaxy and add it to the simulated shear following the linear approximation in Eq.~\eqref{eq:ellipticity_definition}. 

To calculate the aperture mass maps in a computationally inexpensive way, we distribute the galaxies on a grid of pixel size $0.\!'6$. For each pixel, we compute the sum of all respective galaxy ellipticities $\sum_i\epsilon_i$ as well as the sum of their squared absolute ellipticities $\sum_i |\epsilon_i|^2$. The distribution of galaxies on a grid slightly shifts their positions, which introduces a small error in the computed quantity $\epsilon_\mathrm{t}$ when the value $|\vtheta_i-\vtheta|$ of Eq.~\eqref{eq:S/N_definition} is comparable to the size of a pixel, but it enables a significantly faster computation\footnote{ This approximation will likely need to be revisited in future analyses with increased accuracy requirement.}. Moreover, since the convolutions arising in both the numerator and denominator of Eq.~\eqref{eq:S/N_definition} are linear in $\epsilon$ and $|\epsilon|^2$, respectively, computing Eq.~\eqref{eq:S/N_definition} for individual galaxies or for whole pixels does not make a difference,   and the latter is more efficient.
Finally, we chose a filter radius of $\theta_\mathrm{ap}=12.\!'5$ in this work, however the choice of this parameter could be revisited in order to optimise the cosmological constraints.

In addition to exhibiting an irregular shape, our KV450-like footprint is affected by  internally masked regions, e.g.~by   the removal of bright foreground stars,  saturated galaxies or satellite tracks.
Each pixel that contains zero galaxies, which is true in particular for masked regions, is treated as masked. For the calculation of the  aperture mass maps, these masked shear pixels are treated as having a shear of 0,  and we subsequently mask pixels in the signal-to-noise maps for which at least 50\% of the aperture is made of masked shear pixels;  masked $S/N$ pixels are assigned the value $-\infty$. As our simulated data exactly traces the KV450 footprint, our cosmological parameter analysis is not affected by the masked regions. 
 
\subsection{ Topological data analysis}
\label{sec:homology}

 This section describes the methods employed to conduct the topological analysis of the aperture mass maps detailed in Sect.~\ref{subsubsec:numerical_Map}, from 
 the mathematical background to a full description of  our numerical implementation.

\begin{figure*}
\centering
\includegraphics[trim={40 15 40 15},clip,width=0.35\textwidth]{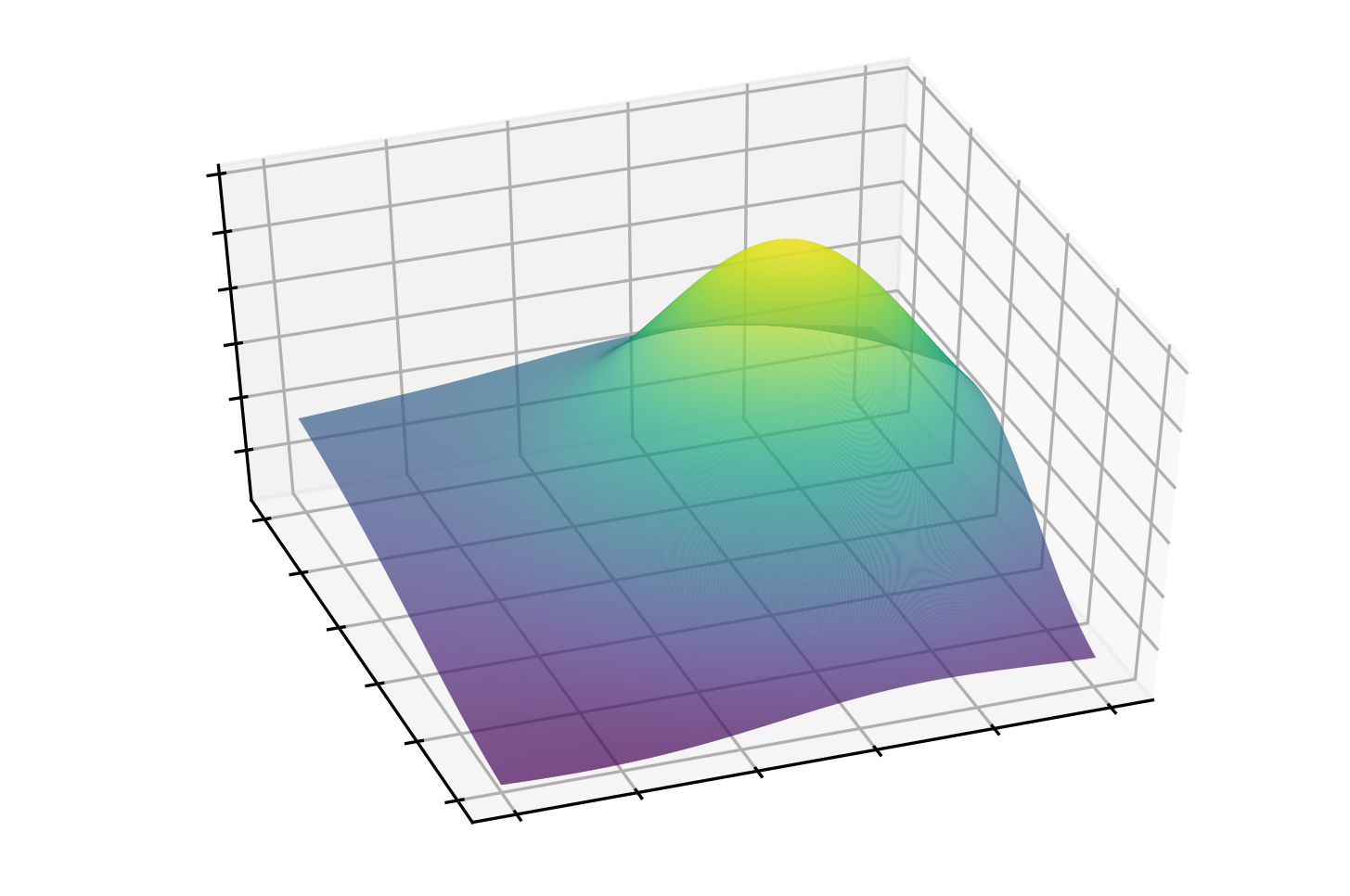}
\includegraphics[trim={40 15 40 15},clip,width=0.35\textwidth]{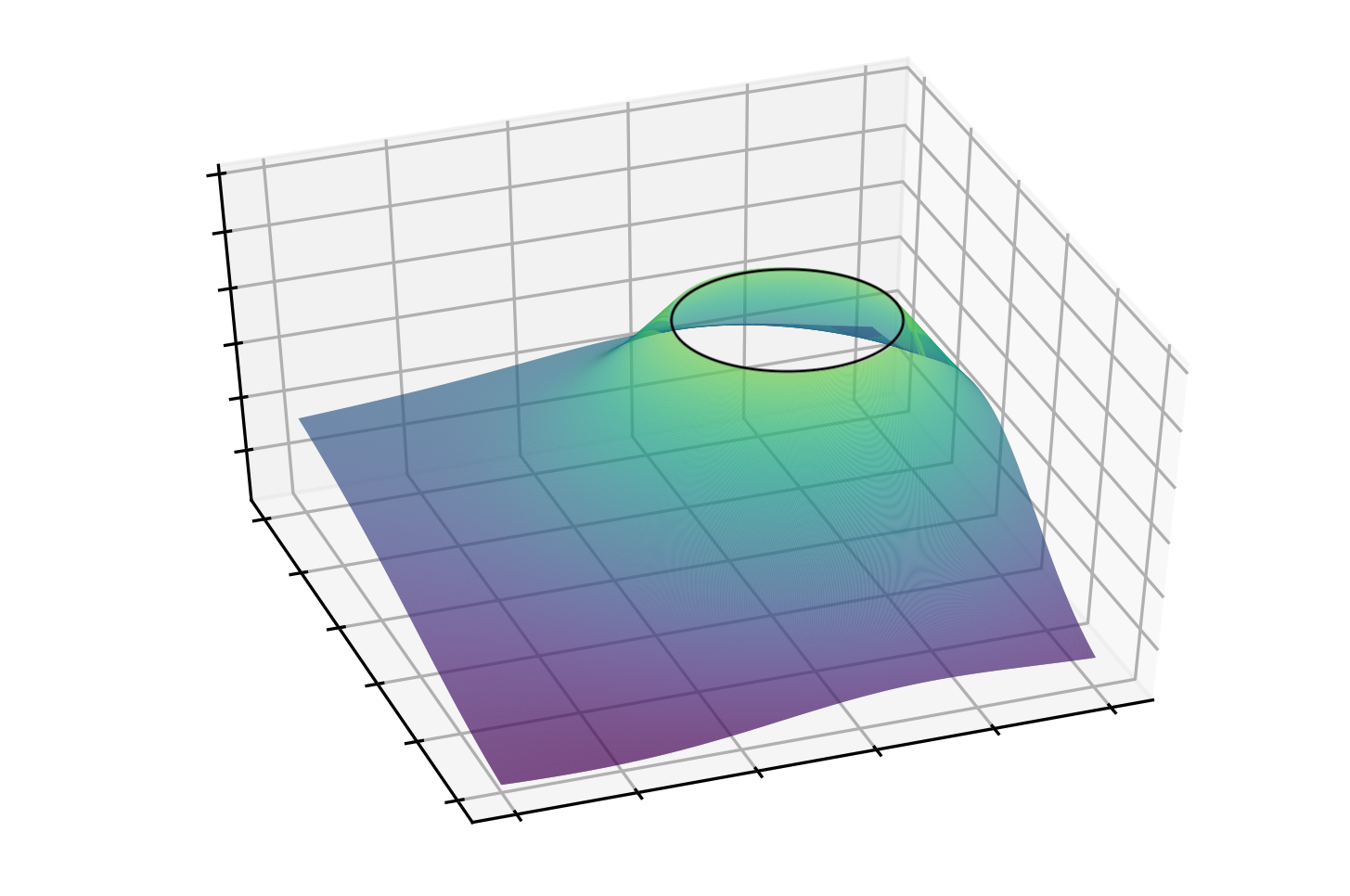}
\mbox{\hspace{0.5cm}\includegraphics[width=0.2\textwidth]{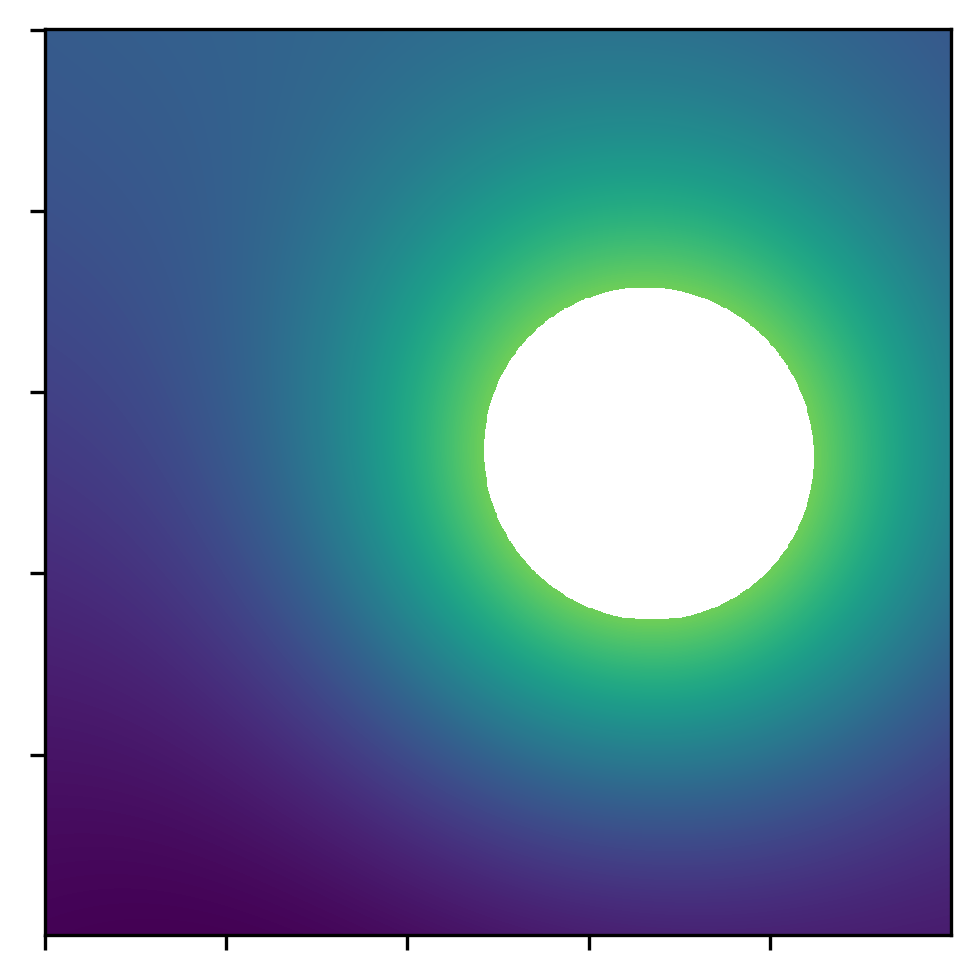}\vspace{1.5cm}}
\caption{Visualisation of an excursion set: the original map (in this case a single peak) is depicted on the left. The middle plot depicts the same peak after a threshold is applied to the map, which cuts off the summit. The result seen from above is depicted on the right, which in this case has one hole (in white) and one connected component (in colour). Varying the threshold generates the full filtration.}
\label{fig:surface_plots}
\end{figure*}

Our analysis is based on the study of topological features in the $S/N$ maps extracted from mock weak lensing data. More particularly, we are interested in studying how these relate to their environment in order to access non-Gaussian information contained in the correlation between different scales of the shear field. 
The idea of our approach is as follows: first, we take the survey area $X$ and we remove from it all pixels that have $S/N$ value above a threshold $t$. The result is a (topological) space that can be seen as a part of $X$ (see Fig.~\ref{fig:surface_plots}). Next, we count the number of connected components\footnote{A connected component is a  cluster of pixels connected by an edge or a corner.} and holes in this space. These are called the {\it Betti numbers} $\betti_0$ and $\betti_1$.
Lastly, we analyse how these numbers change as we vary the threshold $t$. 

Fig.~\ref{fig:visualise_persistence} presents an example of a sequence of spaces obtained in this way. It shows the filtered aperture mass maps for nine values of $t$, obtained from a square-shaped zoom-in region of the full survey area $X$. Inspecting the panels from top-left to bottom-right, we see the gradual recovery of the full aperture mass map as the threshold $t$ increases, starting from the lower values.
One sees that a local minimum of the $S/N$ map corresponds to a connected component appearing at some point and vanishing later on as it gets absorbed by an ``older'' feature; a maximum corresponds to a hole that first appears and later gets filled in.  
Analysing how these features are created and vanish again allows us not only to count the corresponding extrema but to take into account their environment as well, thus obtaining information about the large-scale structure.

\subsubsection{Homology of excursion sets and local extrema}

Mathematically, the ideas described above can be expressed as follows: let $f: X \to \mathbb{R}$ be a map from a topological space $X$ to the reals. In our context, $X$ will be the observed survey area, which we interpret as a subspace of the celestial sphere $S^2$, and $f$ will be the signal-to-noise ratio of the aperture mass, $f(\vtheta) = S/N(\vtheta)$, defined in Eq.~\eqref{eq:S/N_definition}. Taking sublevel sets of the map $f$ (i.e.~the portion of $f$ with values less than some maximal value $t$)  yields a sequence of subspaces of $X$: For $t\in \mbR$, define the \emph{excursion set} $X_t = \set{x \in X \mid f(x) \leq t}$.

To simplify notation, assume that $f$ is bounded. If we choose $t_1 \leq \ldots \leq t_k > \sup(f)$, we obtain a \emph{filtration} of the space $X$, i.e.~a sequence of subspaces with:
\begin{equation}
X_{1} \subseteq X_{2} \subseteq \cdots \subseteq X_{k}=X \, 
\label{eq:inclusion_set}
\end{equation}
where we write $X_i = X_{t_i}$ in the above expression to make the notation more compact.
Our aim is to extract cosmological information\footnote{Note that the theory  described in this section works exactly the same e.g.~for analysing a grey-shade picture where $X$ would be a rectangle and $f$ the intensity at each point; a particular strength of the theory is that it also easily generalises to higher-dimensional data.}
 from the distribution of aperture mass (represented by the map $f$) by studying topological invariants of the sequence $\mathbb{X} = (X_i)_{i}$. 
 
\begin{figure*}
  \centering
  \sidecaption
  \includegraphics[width=\linewidth]{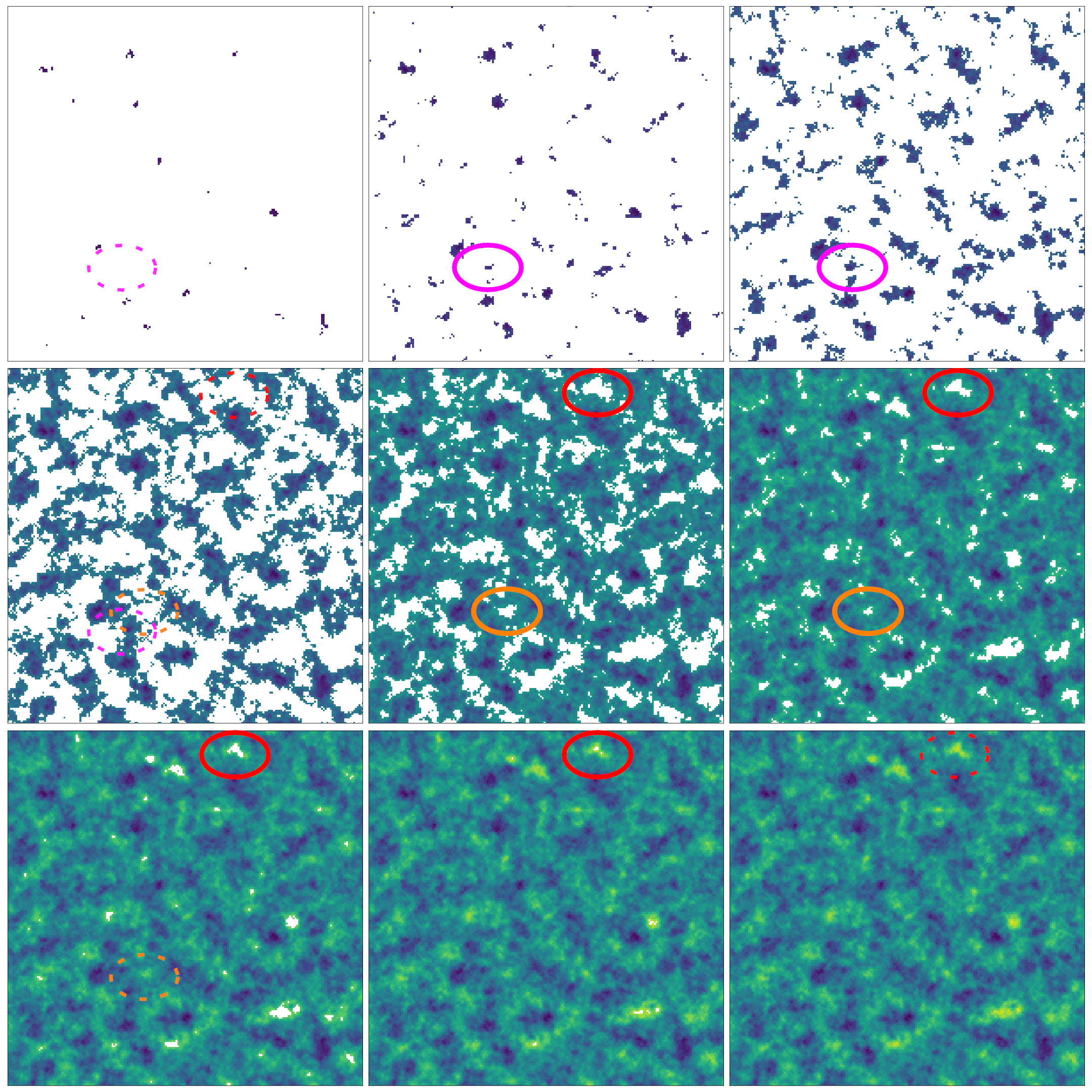}
  \caption{The excursion sets of a sample signal-to-noise map for a $3.3\times 3.3\,\mathrm{deg}^2$ field of SLICS for nine values $t_1, \ldots , t_9$. The solid magenta ellipse highlights a connected component corresponding to a minimum of the signal-to-noise map; the solid orange and red ellipses each highlight a hole corresponding to a maximum. The dashed ellipses indicate the position of the features before their birth and after their death. The magenta feature is born at $t_2$ and dies at $t_4$ after being absorbed by ``older'' features that were born in $t_1$. This lifeline is described by the interval $[t_2,t_4)$ in $\dgm(H_0(\mbX))$. Both the red and orange features are born at $t_5$. The orange one dies at $t_7$, while the red one, which corresponds to a sharper peak, persists longer and dies at $t_9$. Hence, they give rise to intervals $[t_5,t_7)$ and $[t_5,t_9)$ in $\dgm(H_1(\mbX))$.}
  \label{fig:visualise_persistence}
\end{figure*}

A \emph{topological invariant} of a space is a mathematical object (e.g.~a number)  that does not change if we continuously perturb it, e.g.~by stretching or bending. The example that is relevant for us are the Betti numbers. As mentioned above, these count the number of connected components and holes of a space. If we compute the Betti numbers of our excursion sets $X_i$, we obtain information about the number of local extrema of the map $f$: the minima correspond to connected components appearing at some point of the filtration $\mathbb{X}$ and vanishing later on; the maxima correspond to holes (see Fig.~\ref{fig:visualise_persistence}). The time it takes for the connected components or holes to vanish again is related to the relative height of the corresponding extremum, and therefore contains information about the environment.

From a more formal point of view, the $n$-th Betti number $\betti_n(Y)$ of a space $Y$ is computed from its $n$-th \emph{homology group} $H_n(Y)$.   Both homology groups and Betti numbers are important and well-studied invariants from algebraic topology. For an introduction to algebraic topology that is geared towards its applications in the sciences, see \citet{Ghr:ElementaryAppliedTopology};  a more rigorous introduction to this area of mathematics can also be found in \citet{Hat:Algebraictopology}. 
In our context, the homology group $H_n(Y)$ is a vector space that we can associate to $Y$ and $\betti_n(Y)$ is its dimension. The important point for us is that Betti numbers and homology groups translate the geometric (or rather topological) problem of ``counting connected components and holes'' into a question about linear algebra that can be efficiently solved by a computer (see Sect.~\ref{subsubsec:numerical_betti}).

 We note that other topological invariants have been successfully used in cosmology, including the Euler characteristic\footnote{If $Y$ is a $d$-dimensional manifold, then $\chi(Y)$ is, up to a scalar factor, given by the top-dimensional Minkowski functional.} $\chi(Y)$.  Following \citet{1981grf..book.....A}, the Euler characteristic can be used to study real-valued random fields such as the cosmic microwave background \citep{1986ApJ...309....1H,2020A&A...633A..71P}. It is given by the alternating sum of the Betti numbers; in particular, if $Y$ is a subset of $\mbR^2$, we have $\chi(Y)=\betti_0(Y) - \betti_1(Y)$. Hence, one can compute the Euler characteristic from the Betti numbers, but the latter contain strictly more information  \citep[see also the discussion in][]{2019MNRAS.485.4167P},  motivating our choice for this current data analysis. Additionally,  \emph{persistent} Betti numbers, which we  describe in the next subsection, provide us with even more information. The information content extracted from different topological methods can be summarised as follows:
\begin{center}
\begin{tikzcd}[column sep=small]
\arrow[dddd,"\text{\small increasing information}" {rotate=90,anchor=south}, {shorten >= 0.5cm}, {shorten <= 0.5cm}] & \\
 & \text{Euler characteristic} \\
 & \text{Betti numbers} \\
  & \text{persistent Betti numbers} \\
  {}
\end{tikzcd}
\end{center}
This is quantified in Sect.~\ref{sec:inference_of_parameters} (see in particular Fig. \ref{fig:mcmc_result}).

\subsubsection{Persistent homology}
\label{sec:pers_homology}
 As discussed in the last section, one approach to studying the topological properties of the sequence of excursion sets $\mathbb{X} = (X_i)_{i}$ is to   analyse  the sequence of Betti numbers $\betti_n(X_i)$, for $n=0,1$.  Algebraically,
 this amounts to studying the sequence of vector spaces $H_n(X_i)$ and computing each of their dimensions individually. However, one can do better: the inclusion maps $X_i \to X_{i+1}$, presented in Eq.~(\ref{eq:inclusion_set}),  induce linear maps $H_n(X_i) \to H_n(X_{i+1})$ and these provide us with additional information.
The sequence of all $H_n(X_i)$ together with the connecting maps form what is called a \emph{persistence module} that we will write as $H_n(\mbX)$,  and which describes the \emph{persistent homology} of $\mathbb{X}$. For an introduction to persistent homology that focuses on its applications, we refer to \citet{arXiv:1506.08903}; a broader overview and further background material can be found in \citet{Oud:Persistencetheory:quiver}.

The data of the persistence module $H_n(\mbX)$ can be summarised in its \emph{persistence diagram} $\dgm(H_n(\mbX))$. This is a collection of half-open intervals $[t_i,t_j)$, where each interval\footnote{ Such a collection of intervals is called a ``diagram'' because it can be visualised as an actual diagram in the plane: For every interval $[t_i,t_j)$, one can draw a point in $\mbR^2$ with coordinates $(t_i,t_j)$. An example of this can be found in Fig.~\ref{fig:persistence_diagram}.} can be interpreted as a feature\footnote{Formally, these features represent basis elements of the vector spaces in the persistence module.} that is ``born'' at level $t_i$ and ``dies'' at level $t_j$.  
For example, an element in $\dgm(H_0(\mbX))$ is an interval of the form $[t_i,t_j)$ and it corresponds to a connected component that appears at level $t_i$ and merges with another component at level $t_j$. Similarly, elements of $\dgm(H_1(\mbX))$ correspond to holes that appear at $t_i$ and get filled at time $t_j$. The filtration sequence presented  in Fig.~\ref{fig:visualise_persistence} exhibits such features, some of which we have highlighted in the panels: the magenta ellipses present a connected component corresponding to a minimum of the signal-to-noise map; the orange and red ellipses each highlight a hole corresponding to a maximum. As the sequence evolves, these features appear and disappear, giving rise to intervals.

 To conduct our analysis, we study the \emph{$n$-th persistent Betti number} (or \emph{rank invariant}) of $\mbX$. This 
 is defined as the number of features that are born before $t$ and die after $t'$, and can be extracted from the persistence diagram $\dgm(H_n(\mbX))$ by
\begin{equation}
\persbetti_n \mathbb{X}(t,t') = \#\set{[t_i,t_j) \in \dgm(H_n(\mbX)) \mid t_i \leq t \leq t' < t_j}\, ,
\label{eq:persistent_betti}
\end{equation}
where, $\#\set{\ldots}$ denotes the number of elements of the set $\set{\ldots}$. We can consider these as functions from a subset of $\mbR^2$ to $\mbR$; if we want to emphasise this point of view, we will also call them \emph{Betti functions}.
Note that when $t'=t$ we recover the non-persistent Betti numbers:
\begin{equation}
\label{eq:pers-reg-Bettis}
\persbetti_n \mathbb{X}(t_i,t_i) = \betti_n(X_i),
\end{equation}
however the persistent Betti numbers contain strictly more information than the sequence of regular Betti numbers. In particular, if one knows all values of $\persbetti_n \mathbb{X}$, one can recover the entire persistence diagram $\dgm(H_n(\mbX))$, while this is not possible using the sequence $(\betti_n(X_i))_i$. 
Intuitively speaking, the persistent Betti functions do not only provide us with information about how many connected components or holes we have in each excursion set $X_i$, but they also tell us how long each such feature persists throughout the filtration. As explained above, the features correspond to local minima and maxima of the $S/N$ map. Knowing about their lifetimes provides information about the relative height of these extrema and about their entanglement: two completely separated peaks can be distinguished from two peaks sitting together on the summit of a region of high $S/N$.

\subsubsection{Masks and relative homology}

As explained  in Sect. \ref{sec:simulations}, our $S/N$ maps  contain masked regions where we do not have sufficient information about the aperture mass (and their pixel values are set to be constant $-\infty$). In order to incorporate these in our analysis, we work with \emph{relative homology}. The idea of using relative homology to study masked data was also used in P+19, where the authors give further interpretations of these groups. Other occurrences of relative homology in the persistent setting can be found in \citet{PGK:Topologicaltrajectoryclustering} and \citet{BB:Relativepersistenthomology}. Cf.~also extended persistence and variants of it \citep[Section 4 of][]{EH:Persistenthomologysurvey,dSMV:Dualitiespersistentcohomology}.

Relative homology is a variant of ordinary homology, which generally speaking can be thought of as an algorithm that takes as an input a space $X$ and a subspace $M\subseteq X$, and gives as an output for each $n$ a vector space $H_n(X,M)$ that describes the features of $X$ that lie {\it outside} of $M$.
In our set up, we consider the masked regions as the subspace $M$ of our field $X$. As the signal-to-noise map takes the constant value $-\infty$ on this subspace, we have $M \subseteq X_i$ for all $i$ and we get a sequence $(H_n(X_i,M))_i$ of relative homology groups that we want to understand.

Just as before, the inclusions $X_i\to X_{i+1}$ make this sequence into a persistence module $H_n(\mbX,M)$. 
In analogy to Eq.~(\ref{eq:persistent_betti}), its Betti numbers are defined via:
\begin{equation}
\relpersbetti_n \mathbb{X}(t,t') = \#\set{[t_i,t_j) \in \dgm(H_n(\mbX,M)) \mid t_i \leq t \leq t' < t_j}.
\end{equation}

Although the definition of these functions uses relative homology, they can be computed using only ordinary homology:
From the long exact sequence for relative homology \citep[p.~117]{Hat:Algebraictopology}, one can deduce the following formula, which expresses these numbers in terms of the Betti numbers of $\mbX$ and of $M$:
\begin{equation}
\label{eq:formula_rel_persistence}
\relpersbetti_n \mathbb{X}(t_i,t_j) = \persbetti_n\mathbb{X}(t_i,t_j) - \persbetti_n \mathbb{X}(-\infty,t_j) + \betti_{n-1}(M) - \persbetti_{n-1} \mathbb{X}(-\infty,t_i).
\end{equation}
In our case, where $X$ is a proper subspace of the two-sphere $S^2$, the only non-zero terms of these invariants are:
\begin{align}
\nonumber
\relpersbetti_0 \mathbb{X}(t_i,t_j) &= \persbetti_0\mathbb{X}(t_i,t_j) - \persbetti_0 \mathbb{X}(-\infty,t_j), \\
\relpersbetti_1 \mathbb{X}(t_i,t_j) &= \persbetti_1\mathbb{X}(t_i,t_j) - \persbetti_1 \mathbb{X}(-\infty,t_j) + \betti_{0}(M) - \persbetti_{0} \mathbb{X}(-\infty,t_i),\nonumber\\
\relpersbetti_2 \mathbb{X}(t_i,t_j) &= \betti_{1}(M) - \persbetti_{1} \mathbb{X}(-\infty,t_i).
\end{align}
These are the quantities that we want to compute,  and the next section details how this is numerically done.

\subsubsection{Computing persistent Homology: Grids and complexes}
\label{subsubsec:numerical_betti}

As explained in Sect. \ref{sec:simulations}, we compute the signal-to-noise maps on a grid with square-shaped pixels. From a mathematical perspective, subdividing $X$ in this way gives it the structure of a finite cubical complex, i.e.~a space that is obtained by glueing together cubes of maximal dimension (in this case squares) along lower-dimensional cubes (in this case edges and vertices).
We obtain a function on this complex by assigning to each square the value of the map on the corresponding pixel and extending this 
using the lower-star filtration; i.e.~an edge or a vertex gets assigned the minimal filtration value of a square it is adjacent to.
By construction,  the map takes a finite number of values $t_1 \leq \ldots \leq t_k$ on this complex,  which   define the filtration $\mbX$.
We compute the persistent Betti numbers of this filtered complex using the \textsc{Cubical Complexes} module of GUDHI \citep{gudhi:CubicalComplex}\footnote{GUDHI is a well-established, open-source program (available in C++ with a python-interface) for topological data analysis. Among many other applications, it can calculate persistent Betti numbers of multi-dimensional fields. The program is publicly available at \mbox{\url{https://github.com/GUDHI/}}}. 
The masked regions form a subcomplex $M$ of $X$ and we compute $\relpersbetti_n \mathbb{X}(t,t')$ using Eq.~(\ref{eq:formula_rel_persistence}). The number $k$ of steps in the filtration we obtain is quite large, of the order of the number of pixels. GUDHI computes the persistent homology using all these filtration steps, however we evaluate the persistent Betti functions only at a few values (see App.~\ref{sec:evaluation_points}).

 With $X$  the full KV450-like footprint, the full corresponding complex is assembled from 17 tiles (see Sect.~\ref{sec:simulations}), which we label $T^1, \ldots, T^{17}$. In the true KV450 survey, these tiles lie adjacent to each other, and large-scale structures extend across the boundaries. 
In the simulations however, each tile is constructed from a semi-independent realisation, with no tile-to-tile correlation. Therefore, we perform the calculation of aperture masses and the extraction of Betti numbers for each tile individually, making sure that the boundary of each tile is contained in the mask. This implies that the Betti functions $\relpersbetti_n \mathbb{X}(t,t')$ of the entire footprint can be computed as the sum
\begin{equation}
\label{eq:betti_tiles}
\relpersbetti_n \mathbb{X}(t,t') = \sum_{i=1}^{17} \relpersbetti_n \mbT^i(t,t').
\end{equation}

The reason for this is that in $X$, any two points lying in different tiles are separated by points in the mask.
Intuitively speaking, this implies that any feature (e.g.~a connected component or a hole) of $X$ that lies outside the masked regions is entirely contained in one of the tiles $T^i$. Hence, counting the features in $X$ is the same as counting them in each tile and summing them up\footnote{More formally, we have $H_n(X,M) \cong \tilde{H}_n(X/M)$ and this quotient space is given by the wedge sum $X/M \cong \bigvee_{i=1}^{17} T^i/M^i$.}.

\begin{figure*}
  \centering
  \includegraphics[width=\textwidth]{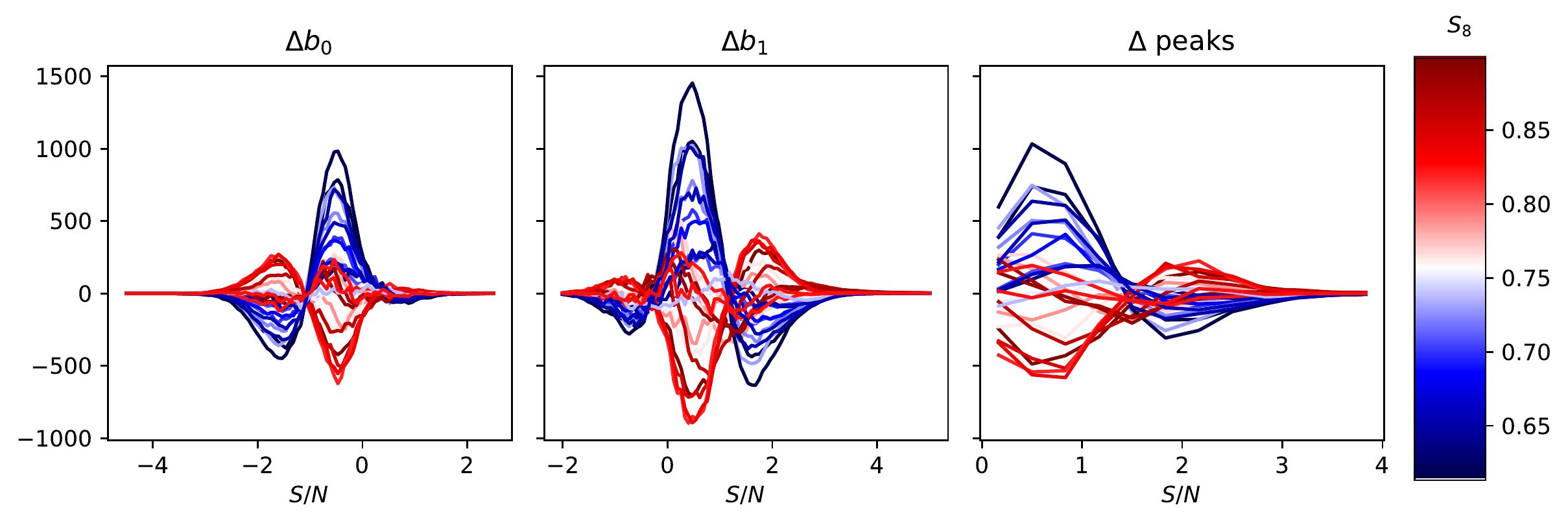}
  \caption{Difference of average Betti numbers $\regbettiKV_{0}$ (left) and $\regbettiKV_{1}$ (center) and peak counts (right) between cosmo-SLICS and SLICS. For the left and the central panels, the $x$-axis represents the respective filtration level $t$, which runs from the minimum to the maximum value of the $S/N$ aperture map. The lines are colour-coded by the value of $S_8$ in the respective cosmo-SLICS (see Tab.~\ref{tab:cosm_param}). The lines shown here are averages over all mock realisations of the full survey footprint (10 for each cosmo-SLICS node, 126 for SLICS).}
  \label{fig:betti_functions}
\end{figure*}

\subsection{Predicting Betti numbers}
\label{sec:gpr}
 As mentioned before, Betti numbers are a highly non-linear statistic, and we were unable to theoretically predict them for a given set of cosmological parameters. Due to the high computational cost of $N$-body simulations, it would also be completely impossible to perform a set of simulations for every point in our given parameter space. Therefore, we chose to emulate the Betti functions for a chosen set of filtration values by computing them in each simulation of cosmo-SLICS, and interpolating the results using the \textsc{Gaussian Process Regression} (GP regression) of scikit-learn \citep{scikit-learn}\footnote{Our algorithm was inspired by the \textsc{GPR Emulator} tool by Benjamin Giblin (\url{https://github.com/benjamingiblin/GPR_Emulator})}.

Simply put, a GP regressor is a machine learning algorithm that takes a training data set $\mathcal{D}$, which consists of $n$ observations of a $d$-dimensional data vector $\vec{y}_i$, along with 
errors in the training data set $\sigma^2(\vec{y})$ and a list of the respective training nodes $\vec{\pi}$. After training, the emulator provides predictions $\vec{y}^*$ for arbitrary coordinates $\vec{\pi}^*$. In our case, the data vectors $\vec{y}_i$ are the Betti functions extracted from the $n=26$ different cosmo-SLICS cosmologies. We set the errors in the training data set $\sigma^2(\vec{y})$ as the variance measured between the different realisations of the  light-cone, which varies with cosmology. Our training nodes $\vec{\pi}$ are the sets of cosmological parameters $\{\Omega_\mathrm{m},\sigma_8,h,w_0 \}$ in each cosmo-SLICS, listed in Table \ref{tab:cosm_param}.
This method offers a model-independent, probabilistic interpolation of multi-dimensional datasets.

 As our kernel, we choose the anisotropic Radial-basis function 
\begin{equation}
  k(\vec{\pi}_i,\vec{\pi}_j) = A\,\exp\left(-\frac{1}{2}d(\vec{\pi}_i/\vec{l},\vec{\pi}_j/\vec{l})^2\right)\; ,
\end{equation}
where $\vec{l}$ is a vector with the same number of dimensions as the input values $\vec{\pi}_i$ (in this case, division is defined element-wise) and $A$ is a scalar. This kernel determines how similar two points $\vec{\pi}_i,\vec{\pi}_j$  are to each other in order to then determine the weights for the individual data points in the interpolation. For each filtration value, we then determine the best hyper-parameters $(A,\vec{l})$ by minimising the log-marginal-likelihood using a gradient descent from 400 different, randomly chosen initial values. For a detailed description of GP regression we refer the interested reader to Appendix A of H+19 and references therein. 

\subsection{Peak count statistics}
\label{subsec:peaks}

 As mentioned in the introduction, we assess the performance of our topological data analysis by comparing its constraining power to that of the peak count statistics, which is another powerful method to capture information of the non-Gaussian part of the matter distribution. This statistic has been increasingly studied in the literature \citep[e.g. M+18,][]{DES-SV-peaks, 2018MNRAS.474.1116S, 2015PhRvD..91f3507L, 2015MNRAS.450.2888L} and is relatively straightforward: it identifies and counts the maxima in the $S/N$ maps, and bins the results as a function of the $S/N$ value at the peak. Peaks of large $S/N$ values convey the majority of the cosmological information (M+18) and correspond to the holes that appear at the latest stages in the filtration sequence presented before. They are typically associated with large galaxy clusters and are less affected by shape noise than peaks of lower $S/N$ values, which however capture additional information from the large-scale structures.
 
In addition to being easy to implement, their constraining power surpasses that of Minkowski functionals \citep{2020arXiv200612506Z} and is competitive with two-point statistics (M+18), making them ideally suited for a performance comparison.

For consistency with the Betti function analysis, we run our peak finder on the exact same $S/N$ maps. For every simulated realisation of the KV450 survey, we count the peaks on the 17 individual tiles and add  the results afterwards. Following M+18, we bin the results in 12 $S/N$ bins ranging from 0.0 to 4.0 and ignore the peaks outside this range. In the right-most panel of  Fig.~\ref{fig:betti_functions}, we present the peak distribution  relative to the mean measurement from the SLICS, colour-coded with the $S_8$ value from the input cosmo-SLICS cosmology. The strong colour gradient illustrates the significant dependence of the signal on $S_8$.

\subsubsection{Comparison between Peaks and Betti numbers}
\label{sec:comp_peaks_Betti}
As explained above, Betti functions are related to the numbers of peaks as well. However, the two methods do not yield completely equivalent information. First, Betti functions do not only take into account the local maxima, but also the minima; this, however, has only a small effect on cosmological parameter constraints since most of the information resides in the peaks (M+18, see also App.~\ref{sec:evaluation_points}). Second, and more importantly, peak count statistics are very local in the sense that they decide about whether a pixel is counted as a peak by simply comparing it to the eight adjacent pixels.

\begin{figure*}
\centering
\includegraphics[width=0.32\textwidth]{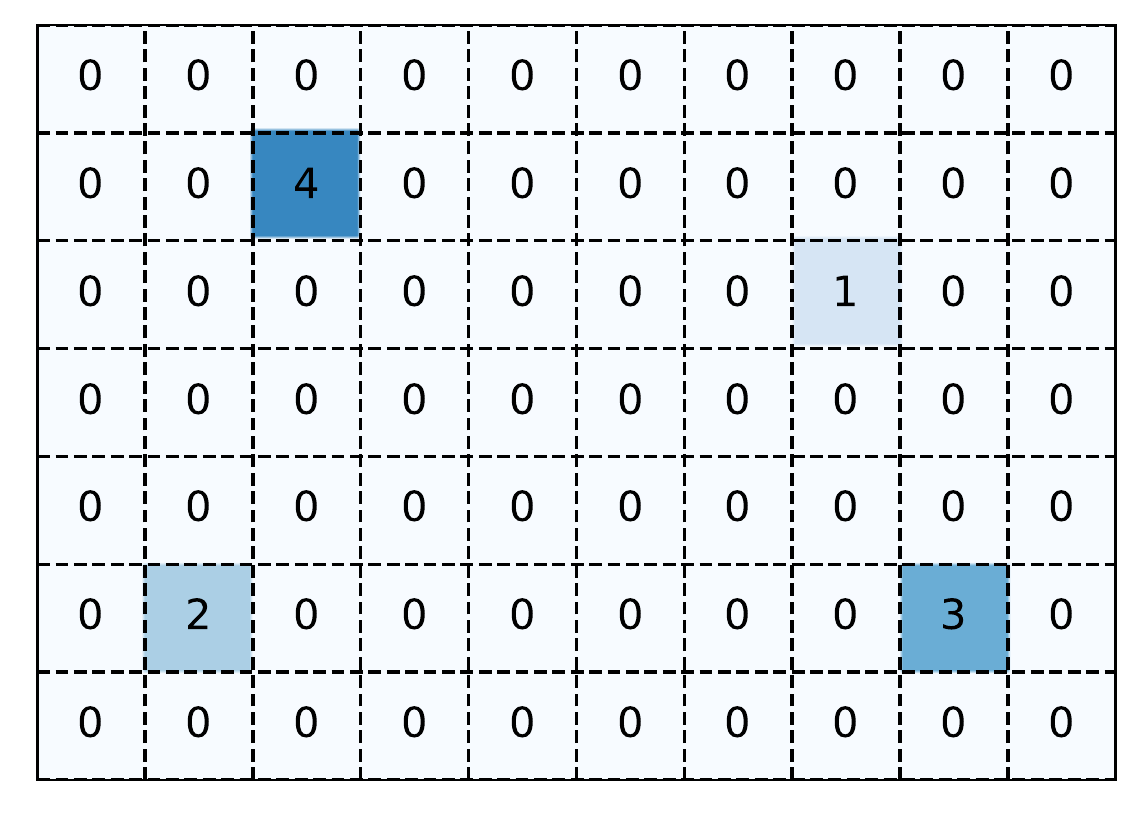}
\includegraphics[width=0.32\textwidth]{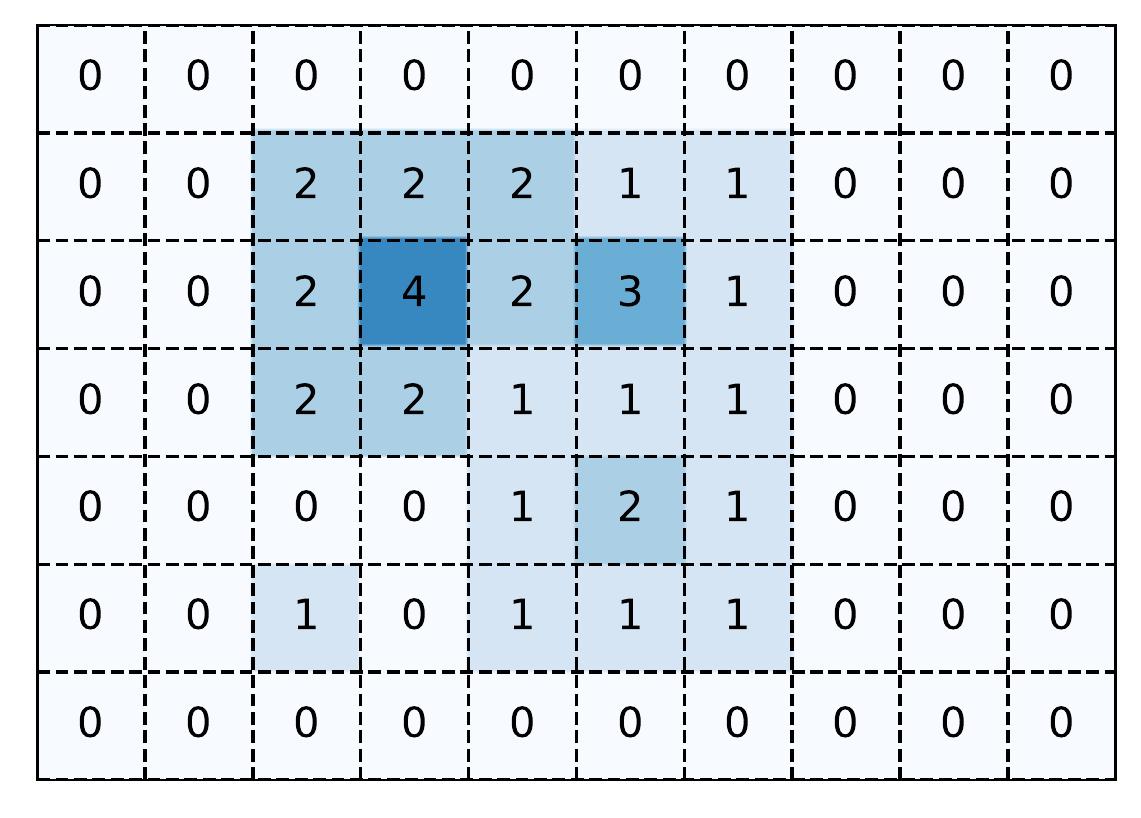}
\includegraphics[width=0.32\textwidth]{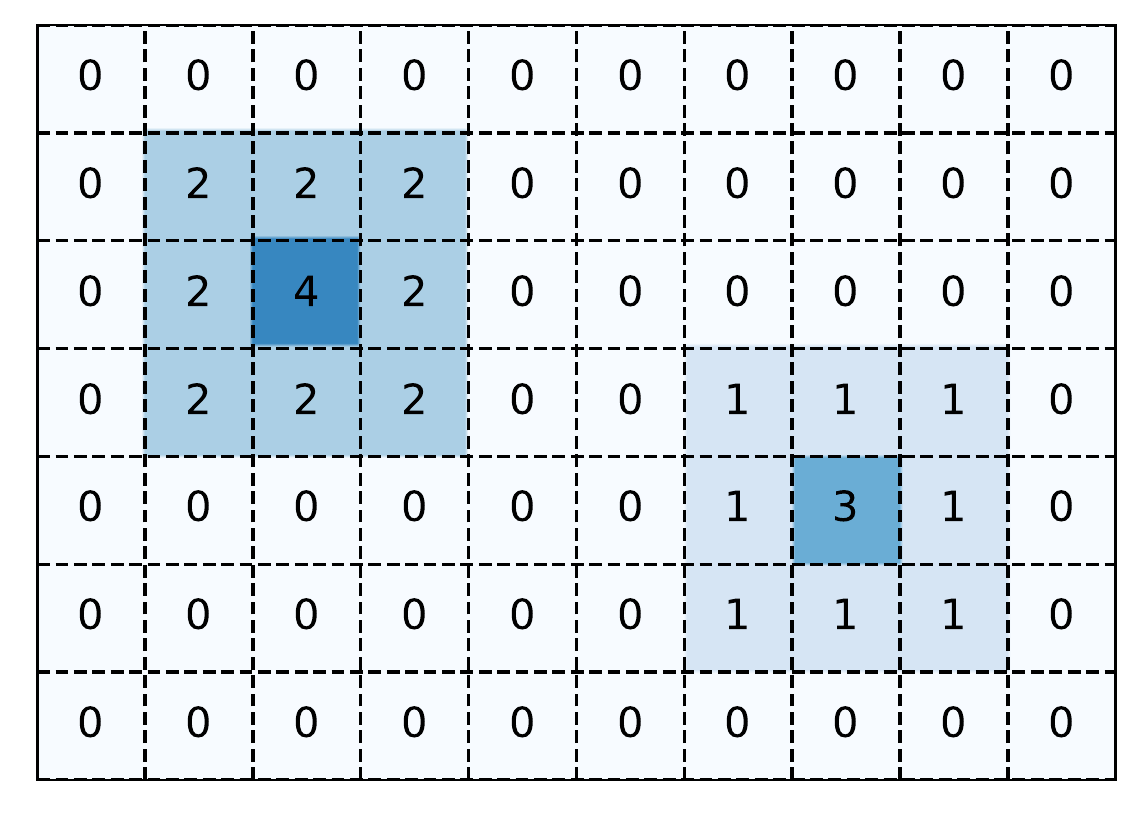}
\caption{Three idealised $S/N$ maps. Both peak count statistics and non-persistent Betti functions can distinguish the first from the third map. However, the peak counts of the first and the second map are identical, as are the non-persistent Betti functions of the second and the third. Persistent Betti functions can distinguish all three maps. For further explanations, see Sect.~\ref{sec:comp_peaks_Betti}.}
\label{fig:comp_peaks_Betti}
\end{figure*}

Fig.~\ref{fig:comp_peaks_Betti} illustrates this with a toy example, presenting three simple cases of reconstructed maps. 
The first two maps cannot be distinguished by peak counts as they both have four peaks with height 1, 2, 3 and 4, respectively.
In contrast, Betti functions are able to identify structures at larger scales such as regions with high $S/N$ values, and are able to differentiate between the first two panels  in Fig.~\ref{fig:comp_peaks_Betti}: In the first map, the excursion set $X_0$ has four holes and these get filled one by one as the threshold $t$ increases. I.e., we have
\begin{equation}
\betti_1(X_t) = \begin{cases}
4 &,0 \leq t <1 ,\\
3 &,1 \leq t <2, \\
2 &,2 \leq t <3, \\
1 &,3 \leq t <4, \\
0 &,4 \leq t. 
\end{cases}
\end{equation}
This is different from the second map, where for all $0\leq t< 3$, the excursion set $X_t$ has only two holes (although the positions of the holes change),
\begin{equation}
\betti_1(X_t) = \begin{cases}
2 &,0 \leq t <3, \\
1 &,3 \leq t <4, \\
0 &,4 \leq t. 
\end{cases}
\end{equation}
However, while the peaks can differentiate between the second and the third map, the sequence of Betti numbers are the same. In the third map as well, we have two holes for all $t$ between 0 and 3 and these get filled at $t=3$ and $t=4$. This illustrates that non-persistent Betti functions cannot distinguish very well between a small number of sharp peaks and a high number of peaks that only slightly protrude from their surroundings. (We ignore $\betti_0$ here as for all $t\geq 0$, all excursion sets have exactly one connected component.)
Persistent Betti functions can differentiate all three cases. They are able to detect that in the second map, the positions of the two holes at $0\leq t < 3$ vary, while in the third map, the same two holes just get filled while $t$ increases. The persistence diagrams $\dgm(H_1(\mbX))$ associated with the three panels are:
\begin{align}
\{ [0,4), [0,3), [0,2), [0,1) \}, && \{ [0,4), [0,1), [1,2), [2,3) \}, \\ 
\nonumber \text{ and } \{ [0,4), [0,3) \}.
\end{align}
As all these are different, so are the associated persistent Betti functions $\persbetti_1 \mathbb{X}$. This simple example illustrates how persistent Betti numbers combine the local information from peak count statistics with the non-local, large-scale information from  non-persistent Betti functions.

\section{Results}
\label{sec:results}
\begin{figure*}
  \centering
  \includegraphics[width=\linewidth]{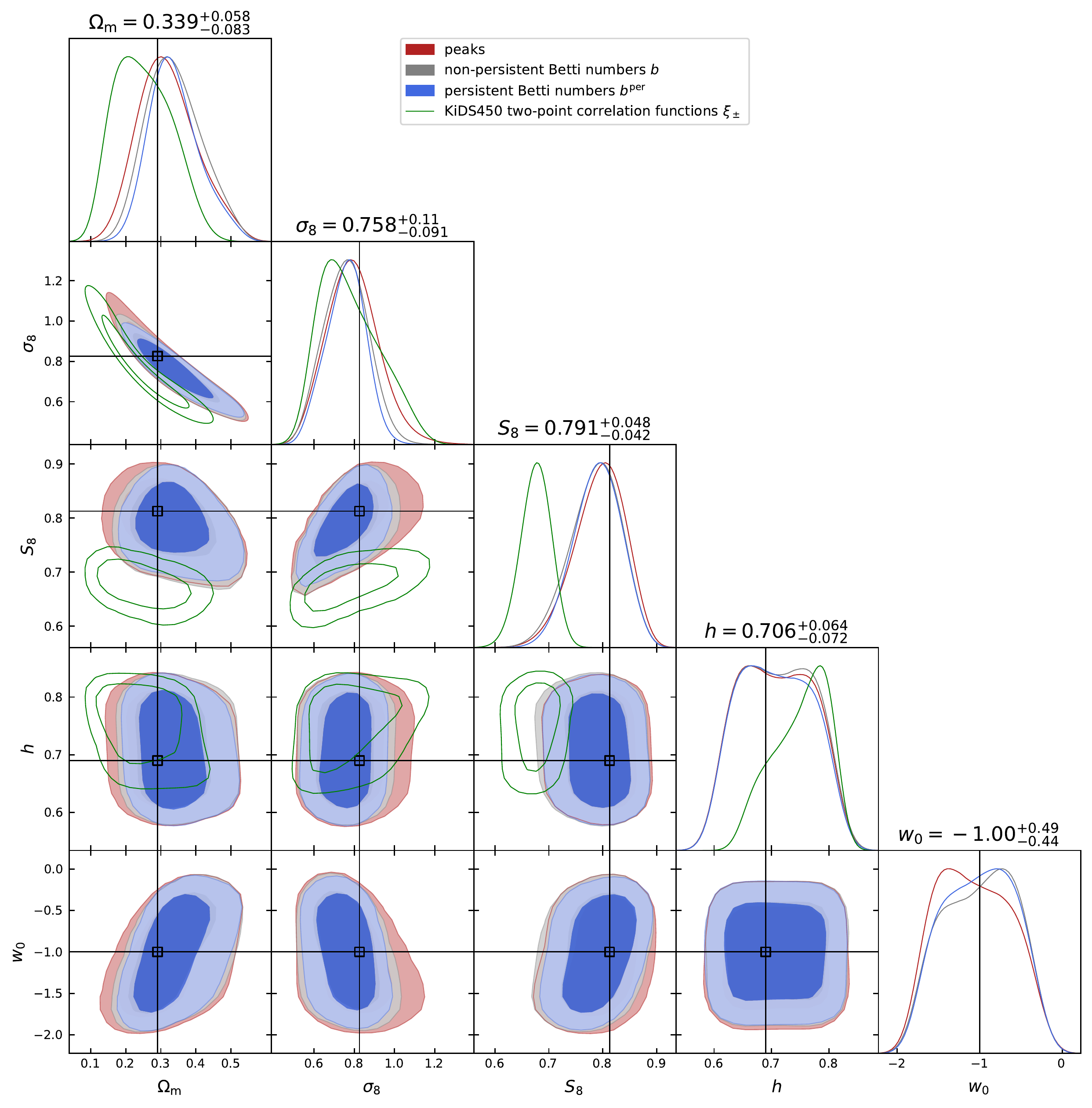}
  \caption{Results of an MCMC on the Betti numbers measured in the KV450-like SLICS. The contours represent the 1- and 2-$\sigma$ posterior contours for an analysis with non-persistent (grey) and persistent (blue) Betti functions. For comparison, the red contours represent the posterior of an analysis with peak statistics, as done in M+18. The values on top of the diagonals represent the marginalised posterior for the persistent Betti functions $\persbettiKV$. The fiducial parameters of SLICS are represented by the black lines. We conducted our analysis with $\Omega_\mathrm{m},\sigma_8,h$ and $w_0$ as free parameters and employed flat priors with $\Omega_\mathrm{m}\in [0.1,0.6]$, $S_8\in [0.6,1.0]$, $h\in [0.5,0.9]$, $w_0\in [-1.8, -0.2]$. The green lines correspond to the 1- and 2-$\sigma$ posterior contours of a cosmological parameter analysis of the KiDS450-survey using tomographic two-point correlation functions, where the covariance matrix was extracted from SLICS and systematics were disregarded \citep[see the ``$N$-body'' setup in Table 4 of][]{2017MNRAS.465.1454H}. Since the set-up of that analysis is very similar to ours, they can be used to compare the relative sizes (but not locations) of contours. Note that this parameter analysis was done in a $\Lambda$CDM-framework on KiDS data, whereas our analysis is done in a $w$CDM framework on mock data.}
  \label{fig:mcmc_result}
\end{figure*}

 In this section, we present the results from our topological analysis based on the persistent Betti numbers. We gauge the performance of the technique from a comparison with an analysis based on their non-persistent alternatives, and additionally with a peak statistics analysis. We first present our measurements from the simulations, which serve to train the emulator and estimate the covariance matrices, then proceed with the parameter inference.

\subsection{Calibrating the emulator and determining the covariances}
\label{sec:covariance_matrix}

For each light-cone of  the cosmo-SLICS simulation suite, we measure  the persistent Betti function $\persbettiKV_n(t,t') = \relpersbetti_n \mathbb{X}(t,t')$, and compute a non-persistent version $\regbettiKV_n$ by setting $\regbettiKV_n(t)=\persbettiKV_n(t,t) = \relpersbetti_n \mathbb{X}(t,t)$.
We extract the average values of  $\regbettiKV_{0,1}$ and $\persbettiKV_{0,1}$ at a chosen set of values for each of the 26 different cosmologies from the mean over the 10 light-cones.

To  demonstrate that Betti functions are indeed sensitive to the underlying cosmological parameters, we show the non-persistent Betti functions\footnote{We focus on the non-persistent case here, since the persistent Betti numbers depend on tuples of evaluation points, which cannot be visualised easily.} $\regbettiKV_n$ for different input cosmologies in Fig.~\ref{fig:betti_functions}, again colour-coded as a function of $S_8$. As for the peak statistics, there is a strong colour gradient at every element of the data vector, both for $\regbettiKV_0$ and $\regbettiKV_1$, which indicates the sensitivity to that parameter.
 
In both the non-persistent and persistent settings, we obtain a data vector consisting of the function values at the chosen evaluation points (introduced in Sect. \ref{subsubsec:numerical_betti} and discussed in App.~\ref{sec:evaluation_points}), and further concatenate these values for $\regbettiKV_0$ and $\regbettiKV_1$ or $\persbettiKV_0$ and $\persbettiKV_1$, respectively. We do not consider $\regbettiKV_2$ and $\persbettiKV_2$ for our analysis, as they only consist of a very small number of features, which mainly depend on the shape of the mask and contain very little cosmological information. We use these data vectors to train the hyperparameters of our  GPR emulator, and are hereafter able to predict the Betti functions (at the chosen evaluation points) for arbitrary cosmological parameters enclosed within the training range.

 As a next step, we extract the same Betti functions for each realisation of the SLICS and use these to determine a covariance matrix $C_\mathrm{b}$ for the  data vector.

To assess the accuracy of the emulator, we perform a  ``leave-one-out cross-validation test'': we exclude one cosmology node and train our emulator on the 25 remaining ones. We then let the emulator predict Betti functions for the previously excluded cosmology and compare it to the measured values; this process is repeated for each of the 26 cosmologies. The results are presented    in App. \ref{sec:emulator_accuracy}. As further discussed therein, this test demonstrates that we achieve an accuracy of a few percent on our predicted signal at most evaluation points. While a single outlier contains inaccuracies that are larger than the $1\sigma$ limit of the sample variance extracted from SLICS, the vast majority of cross-validation tests is contained within this limit. For each evaluation point of the Betti functions, we further estimate the variance by
\begin{equation}
 \sigma^2_{\regbettiKV_n} = \frac{1}{25}\sum_{i=0}^{25}\left( x_{i,\mathrm{measured}} - x_{i,\mathrm{predicted}}\right)^2\; ,
\label{eq:emulator_variance}
\end{equation}
where $x_{i,\mathrm{measured}}$ is the measured value of the Betti function at the $i$-th training cosmology, whereas $x_{i,\mathrm{predicted}}$ is the one predicted by the emulator when it is trained on all other cosmologies, leaving the $i$-th one  out. We define the ``covariance matrix'' of the emulator $C_\mathrm{e}$ as a diagonal matrix  where the entries correspond to the variance of the respective filtration value, given by Eq. \eqref{eq:emulator_variance}. Combining with the sample variance $C_\mathrm{b}$ estimated from the SLICS, we subsequently set 
\begin{equation}
C=C_\mathrm{b}+C_\mathrm{e}
\label{eq:covmat_addition}
\end{equation}
as the covariance matrix for our cosmological parameter estimation. We note that  $C_\mathrm{e}$ is likely to overpredict the true emulator uncertainties: this is because the training nodes that reside in regions close to the limits of our parameter space contribute to the majority of the error budget, whereas the majority of evaluations in a cosmological parameter analysis is expected to be in densely sampled regions.

 We finally repeat this full machinery (GPR training and covariance estimation), this time on the peak count statistics measurements extracted from the same data sets and described in Sect.~\ref{subsec:peaks}; we compare the respective performances in the next section.

\subsection{Inference of cosmological parameters}
\label{sec:inference_of_parameters}

 Putting to use the covariance matrix and the prediction tools described in the last section, we perform here a trial cosmological parameter inference using the mean of the SLICS measurements as our ``observed data'' to test how well we can recover the fiducial values, and how the constraining power on the cosmological parameters compares with the peak count statistics described in Sect.~\ref{subsec:peaks}. We sample our cosmological parameter space using the MCMC sampler \textsc{emcee} \citep{arXiv:1202.3665}, specifying a flat prior range that reflects the parameter space sampled by the cosmo-SLICS, namely: $\Omega_\mathrm{m}\in [0.1,0.6]$, $S_8\in [0.6,1.0]$, $h\in [0.5,0.9]$, $w_0\in [-1.8, -0.2]$. 
Since our covariance matrix is extracted from a finite number of simulations, it is inherently noisy, which leads to a biased inverse covariance matrix \citep{Hartlap:2007}. To mitigate this, we adopt a multivariate $t$-distribution \citep{Sellentin:2016} as our likelihood model. At each sampled point, we compute the log-likelihood as 
\begin{equation}
  \log(\mathcal{L}) = -\frac{1}{2} N_\mathrm{s} \log\left(1+\frac{\chi^2}{N_\mathrm{s}-1}\right)\, ,
  \label{eq:loglikelihood}
\end{equation}
where $N_\mathrm{s}$ is the number of simulations used to calibrate the covariance matrix (in our case 126), and $\chi^2$ is computed in the usual way via
\begin{equation}
\chi^2 = (\vec{x}_\mathrm{measured}-\vec{x}_\mathrm{predicted})^T \, C^{-1} \, (\vec{x}_\mathrm{measured}-\vec{x}_\mathrm{predicted}) \; .
\end{equation}
Here, $\vec{x}_\mathrm{measured}$ are the respective Betti functions $\regbettiKV_{0,1}$ or $\persbettiKV_{0,1}$ extracted from SLICS, while $\vec{x}_\mathrm{predicted}$ are predicted by the emulator at the respective point in parameter space.

Following these steps, we perform an MCMC analysis  both for the persistent and the non-persistent Betti functions, $\persbettiKV_{0,1}$ and $\regbettiKV_{0,1}$ and for the peak count statistics. The results are displayed in Fig.~\ref{fig:mcmc_result}. 
As can be seen, all methods recover the fiducial simulation parameter within the $1\sigma$ limit. Furthermore, the persistent Betti functions $\persbettiKV$ offer slightly improved constraining power when compared to the non-persistent version $\regbettiKV$  and to peak counts, and is notably the best at rejecting the low-$\Omega_{\rm m}$ and the high-$\sigma_8$ regions of the parameter space. The $1\sigma$ contours are all smaller, yielding marginalised constraints on $S_8$ of $0.791^{+0.048}_{-0.042}$, compared to $0.787^{+0.053}_{-0.042}$ for non-persistent Betti numbers and  $0.795^{+0.053}_{-0.042}$ for peaks. Persistent homology therefore increases the constraining power on $S_8$ by 5\% compared to peak statistics, while the ``figure of merit'' (i.e.~one over the area of the $1\sigma$ contour) in the $\Omega_\mathrm{m}-\sigma_8$ plane is increased by 44\%.  This is not entirely surprising, as $b$ and peaks just counts features at each filtration step, whereas $\persbettiKV$ contains the entire information content of the persistence module, as discussed in Sect.~\ref{sec:pers_homology}. 

For comparison, we also report in Fig.~\ref{fig:mcmc_result} the constraints from a two-point correlation function analysis (see the green curves). These were obtained from the actual measurements from the KiDS-450 data, which had a slightly different redshift distribution and number density  than the KV450-like data used in this paper, but were derived with a covariance matrix extracted from the SLICS and ignored systematic uncertainty  \citep[see the ``$N$-body'' setup in Table 4 of][]{2017MNRAS.465.1454H}, in a setup otherwise very similar to ours. The contours are offset to lower $S_8$ values as preferred by the KiDS-450 data, in contrast with the higher input SLICS cosmology. Due to the very similar analysis setup, the contours can be used to compare their size with the ones derived from our analysis. The increased constraining power with respect to peaks and homology is largely due to the tomographic analysis, which we reserve for future work.

It is worth noting that a part of the reported uncertainties for persistent Betti functions arise due to the inaccuracy of the emulator that is larger than for peaks; when these uncertainties are ignored, persistent Betti numbers achieve constraints of $S_8 = 0.786^{+0.045}_{-0.040}$, an improvement of 8\% with respect to peak statistics. This can potentially be realised in a future analysis involving a larger training set (we further discuss this in App.~\ref{sec:emulator_accuracy}.) Although a gain of 5-8\% in accuracy on $S_8$  is small in comparison with the infrastructure work required, we show in the next section that the improvement is significantly increased once the weak lensing data better resolves the large scale structure around the peaks.

Before steering away from the KV450-like setup however, we finally investigate the systematic impact of two other survey parameters. Similarly to the shear peak count statistics, the Betti functions are extracted from the signal-to-noise ratio of the aperture mass map. Therefore, their expectation value directly depends not only on the survey footprint but also on various terms influencing the noise, for example the effective number density of galaxies and their average internal ellipticity (see e.g.~M+18). For the two-point correlation functions, these terms only influence the covariance matrix, at least to first order \citep{2020A&A...634A.104H}, while here a 10\% offset on the galaxy density or on the shape noise can bias the inferred parameters by 1-3$\sigma$.  Full details of these tests are provided  in App.~\ref{sec:systematics}.

\section{Outlook for Stage IV experiments}
\label{sec:euclid}

As discussed in Sect.~\ref{subsec:peaks}, the gain of information from persistent Betti functions with respect to peak counting comes from the ability to extract information about the environment, e.g.~by distinguishing whether peaks are isolated or clustered on a larger over-density. Consequently, we expect that resolving the large-scale structure with  higher resolution, as will be made possible with upcoming Stage IV lensing experiments, should further increase this relative gain.

In this section, we explore this scenario by repeating the analysis presented previously on a  set of simulations with \textit{Euclid}-like source number densities. These are also constructed from the SLICS and cosmo-SLICS light-cones, but differ in a few key aspects: in contrast to the KV450-like mock data, the position of the galaxies are here placed at random on the  $10\times 10\,\mathrm{deg}^2$ fields, and no masks are imposed on them. Our total survey area is therefore 100 deg$^2$. The redshift distribution follows: 
\begin{equation}
n(z) \propto z^{2} {\rm exp} \Bigg[-\Bigg(\frac{z}{z_0}\Bigg)^{\beta}  \Bigg]
\end{equation}
with $z_0 = 0.637$, $\beta=1.5$, and the overall proportionality constant is given by normalising the distribution to 30 gal/arcmin$^2$. 

In this analysis, we opted for an aggressive strategy in which we include peaks and features with $S/N$ up to 10, which might end up being rejected in the future due to difficulties at modelling all systematic effects (e.g.~baryons feedback) at the required level. We carried out a second analysis which cap the features at $S/N$ of 7 instead, and noticed only minor differences. In both cases, the peak statistics were binned such that a signal-to-noise range of 1 is covered by 3 bins.

One important aspect to note is that due to the much lower level of shape-noise in this Stage IV setup, the accuracy of the GPR emulator degrades in comparison to the KV450-like scenario, reaching $\sim\! 5\%$ only for a few evaluation points.
More details about the relative importance of the emulator accuracy are provided in App.~\ref{sec:emulator_accuracy}.
Meanwhile, we have included the error associated with the emulator following Eqs.~\eqref{eq:emulator_variance} and \eqref{eq:covmat_addition} in our analysis, hence our results should remain unbiased. 

\begin{figure*}
  \centering
  \includegraphics[width=\linewidth]{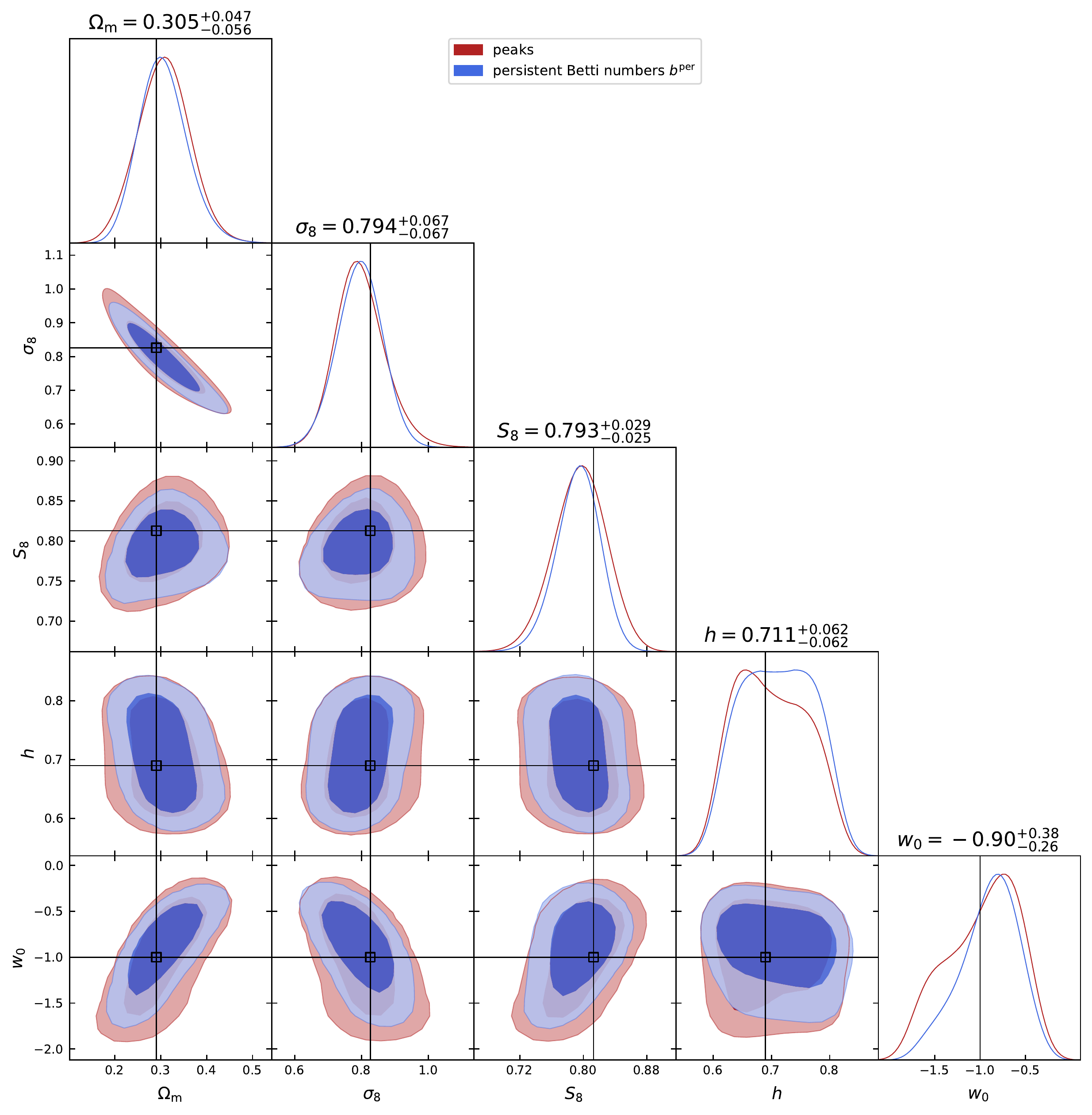}
  \caption{Results of an MCMC as described in Fig. \ref{fig:mcmc_result}, but here for 100 deg$^2$ of \textit{Euclid}-like lensing data.}
  \label{fig:mcmc_result_euclid}
\end{figure*}
 
The results of our MCMC analysis are shown in Fig.~\ref{fig:mcmc_result_euclid}, where we compare the performance of the persistent Betti numbers to peak counting. We find here again an increase in statistical power and the gain is amplified compared with the KV450-like analysis: looking at  the $S_8$ constraints, we measure $S_8 = 0.796^{+0.033}_{-0.033}$ for peak statistics and $S_8 = 0.792^{+0.029}_{-0.025}$ for persistent Betti numbers, an improvement of $\sim\! 18\%$. Contraints on $\Omega_{\rm m}$ are also improved, with a gain of $\sim\!10\%$ on the one-dimensional marginal error. Persistent Betti numbers are even able to set some constraints on the dark energy equation of state, measuring $w_0=-0.90^{+0.38}_{-0.26}$, whereas the constraints from peaks are completely given by the prior. This is an exciting new avenue that will be further investigated in future work involving tomographic decomposition of the source catalogues.

We emphasise that for the persistent Betti numbers, the emulator inaccuracy constitutes a large part of the error budget. This uncertainty could be reduced by increasing the number of training nodes (see App.~\ref{sec:emulator_accuracy}). When disregarding the emulator uncertainty, we achieve constraints of $S_8 = 0.793^{+0.026}_{-0.023}$ and $w_0=-0.91^{+0.29}_{-0.19}$.

\section{Discussion}
\label{sec:discussion}
 Based on the topological analysis of an ensemble of  realistic numerical simulations, we have demonstrated that persistent Betti functions are a highly competitive method to constrain cosmological parameters from weak lensing data. We carried out an MCMC sampling and found that persistent Betti functions improve the constraining power on $S_8$ by $\sim\! 5$-$8\%$ compared  to peak count statistics, which themselves have a constraining power that is similar to the mainstream two-point statistics (M+18). Furthermore, this advantage over peak statistics is expected to exceed 20\% in the upcoming Stage IV surveys. We include the effect of shape noise and of masking as they occur within the KV450 weak lensing survey, however, our methods can be directly adapted to other surveys. 

Our results are obtained by estimating a covariance matrix from an ensemble of independent light cones, and by training a Gaussian Process Regression emulator on a suite of $w$CDM simulations. A non-negligible part of our error budget arises from inaccuracies in the emulator. These occur since we can only calibrate our model for 26 different cosmologies, which can certainly be improved in the future.

We investigated the influence of shape noise and number density on our analysis (see App.~\ref{sec:systematics}). However, various other systematic effects inherent to weak lensing data analyses were not explored in this paper, including the uncertainty arising from photometric redshift errors, shape calibration, baryon feedback on the matter distribution or intrinsic alignments of galaxies. Before a full cosmological parameter analysis can be carried out from data, it is crucial to investigate and understand how to include these. Additionally, there are many  internal parameters in our analysis pipeline that  were chosen by hand, for example the aperture radius $\theta_\mathrm{ap}$ of the filter function or the points at which we evaluate the Betti functions. With careful optimisation, one would probably be able to extract an even higher amount of information from the persistent Betti numbers.

 We also want to point out that persistent Betti numbers are just one way to compress the information of persistent homology. While they are certainly easy to understand and apply, they also suffer from some disadvantages, the most notable  of which being that the difference between two Betti functions is always integer-valued, whereas we would prefer a real-valued distance function for a $\chi^2$-analysis. This problem can be mitigated by utilising different statistics of persistent homology  \citep[e.g.][]{RHBK:stablemultiscale,Bub:Statisticaltopologicaldata}; for an overview of further options, see also \citet[Chapter 8]{Oud:Persistencetheory:quiver} and \citet{PXL:PersistentHomologybased}.

Lastly, we note that  topological data analysis is a promising avenue in cosmology that can find multiple applications well outside those presented in this paper. Other methods similar or complementary to persistent Betti functions can be used to detect and quantify structure in both continuous fields and point-clouds in arbitrary dimensions \citep[see e.g.][and references therein]{arXiv:1506.08903}, which makes topological data analysis an incredibly versatile tool to study the distribution of matter in our Universe. While it has been used in cosmology before, e.g.~to detect non-Gaussianities in the cosmic microwave background (see P+19) or to find voids in the large-scale structure \citep{2019A&C....27...34X}, its utilisation in modern astronomy is bound to grow in the near future.
\begin{acknowledgements}
We are grateful for useful hints from Fabian Lenzen and Vidit Nanda on persistent homology and from Pierre Burger and Peter Schneider on the data analysis. Furthermore we thank Shahab Joudaki, Konrad Kuijken, Frank R\"ottger and Sandra Unruh for helpful comments on previous versions of this article.\\

Sven Heydenreich acknowledges support from the German Research Foundation (DFG SCHN 342/13) and the International Max-Planck Research School (IMPRS).
Benjamin Br\"uck was supported by the Danish National Research Foundation through the Copenhagen Centre for Geometry and Topology (DNRF151).
Joachim Harnois-D\'eraps acknowledges support from an STFC Ernest Rutherford Fellowship (project reference ST/S004858/1). Computations for the $N$-body simulations were enabled by Compute Ontario (www.computeontario.ca), Westgrid (www.westgrid.ca) and Compute Canada (www.computecanada.ca). The SLICS numerical simulations can be found at http://slics.roe.ac.uk/, while the cosmo-SLICS can be made available upon request.\\
\\
\emph{Author contributions}: All authors contributed to the development and writing of this paper. SH led the data analysis, BB developed the necessary mathematical background, while JHD provided the suites of numerical simulation tailored to the measurement.

\end{acknowledgements}

\bibliographystyle{aa}
\bibliography{cite,non_ads_refs}

\begin{appendix} 
\section{Additional information about the cosmo-SLICS simulations}

 For completeness, we list in Table \ref{tab:cosm_param} the full suite of cosmological parameters that makes up the cosmo-SLICS series. Full details about this simulation suite can be found in H+19.

\begin{table}[h]
\caption{Cosmological parameters of the cosmo-SLICS $w$CDM simulations. Other fixed parameters are $\Omega_\mathrm{b} = 0.0447$ and  $n_s = 0.969$.}
\label{tab:cosm_param}
\centering
\begin{tabular}{c | c c c c c c }
 & $\Omega_\mathrm{m}$ & $S_8$ & $h$ & $w_0$ & $\sigma_8$ & $\Omega_\mathrm{cdm}$ \\
\hline
\hline
00 & 0.3282 & 0.6984 & 0.6766 & -1.2376 & 0.6677 & 0.2809  \\
01 & 0.1019 & 0.7826 & 0.7104 & -1.6154 & 1.3428 & 0.0546  \\
02 & 0.2536 & 0.6133 & 0.6238 & -1.7698 & 0.667 & 0.2063  \\
03 & 0.1734 & 0.7284 & 0.6584 & -0.5223 & 0.9581 & 0.1261  \\
04 & 0.3759 & 0.8986 & 0.6034 & -0.9741 & 0.8028 & 0.3286  \\
05 & 0.4758 & 0.7618 & 0.7459 & -1.3046 & 0.6049 & 0.4285  \\
06 & 0.1458 & 0.768 & 0.8031 & -1.4498 & 1.1017 & 0.0985  \\
07 & 0.3099 & 0.7861 & 0.694 & -1.8784 & 0.7734 & 0.2626  \\
08 & 0.4815 & 0.6804 & 0.6374 & -0.7737 & 0.5371 & 0.4342  \\
09 & 0.3425 & 0.7054 & 0.8006 & -1.501 & 0.6602 & 0.2952  \\
10 & 0.5482 & 0.6375 & 0.7645 & -1.9127 & 0.4716 & 0.5009  \\
11 & 0.2898 & 0.7218 & 0.6505 & -0.6649 & 0.7344 & 0.2425  \\
12 & 0.4247 & 0.7511 & 0.6819 & -1.1986 & 0.6313 & 0.3774  \\
13 & 0.3979 & 0.8476 & 0.7833 & -1.1088 & 0.736 & 0.3506  \\
14 & 0.1691 & 0.8618 & 0.789 & -1.6903 & 1.1479 & 0.1218  \\
15 & 0.1255 & 0.6131 & 0.7567 & -0.9878 & 0.9479 & 0.0782  \\
16 & 0.5148 & 0.8178 & 0.6691 & -1.3812 & 0.6243 & 0.4675  \\
17 & 0.1928 & 0.8862 & 0.6285 & -0.8564 & 1.1055 & 0.1455  \\
18 & 0.2784 & 0.65 & 0.7151 & -1.0673 & 0.6747 & 0.2311  \\
19 & 0.2106 & 0.8759 & 0.7388 & -0.5667 & 1.0454 & 0.1633 \\
20 & 0.443 & 0.8356 & 0.6161 & -1.7037 & 0.6876 & 0.3957  \\
21 & 0.4062 & 0.662 & 0.8129 & -1.9866 & 0.5689 & 0.3589  \\
22 & 0.2294 & 0.8226 & 0.7706 & -0.8602 & 0.9407 & 0.1821  \\
23 & 0.5095 & 0.7366 & 0.6988 & -0.7164 & 0.5652 & 0.4622  \\
24 & 0.3652 & 0.6574 & 0.7271 & -1.5414 & 0.5958 & 0.3179  \\
fid & 0.2905 & 0.8231 & 0.6898 & -1.0 & 0.8364 & 0.2432  \\
\end{tabular}
\end{table}

\section{Choosing evaluation points for the Betti functions}
\label{sec:evaluation_points}
While the domain of both persistent and non-persistent Betti functions is technically limited by the pixel-values of the signal-to-noise map, the  raw data vector provided by GUDHI contains about $10^6$ entries. As this level of refinement is neither practical nor necessary, we need to choose points at which to evaluate and compare the Betti functions. 
To help in this decision, we compute the persistent Betti numbers for SLICS and cosmo-SLICS on a dense grid. We then compare the mean squared differences between SLICS and cosmo-SLICS, weighted by the inverse variance of the respective Betti numbers within SLICS. This basically measures how well an evaluation point can distinguish different cosmologies. We rank the evaluation points according to this discrimination potential, starting with the ``best'' evaluation point. We then recursively build our set of evaluation points in the following way:

Given a set of evaluation points, we extract the mean of the persistent Betti numbers measured from SLICS, $\vec{x}_\mathrm{SLICS}$, and the same data vector for each cosmology $i$ of the cosmo-SLICS, $\vec{x}_i$. Furthermore, we extract a covariance matrix C for this data vector from SLICS. Using these, we compute the quantity $\tilde{\chi}^2$ via:
\begin{equation}
\tilde{\chi}^2 = \frac{1}{26} \sum_i\left[(\vec{x}_\mathrm{SLICS}-\vec{x}_i)^TC^{-1}(\vec{x}_\mathrm{SLICS}-\vec{x}_i)\right] \, .
\end{equation}
We then add the next evaluation point to our data vector and check whether the new $\tilde{\chi}^2$ surpasses the old one by a certain threshold (in our case 0.2), while additionally demanding a minimum of 40 features per line-of-sight\footnote{This ensures that we avoid tuning our analysis on artefacts that are dominated by shot-noise.}. If those two criteria are fulfilled, we add this evaluation point to the data vector and repeat the process. If not, we check the next-best evaluation point. This yields a total of 33 evaluation points for the KV450-like survey, and 46 evaluation points for the {\it Euclid}-like survey. We note that this method favours $b_1$ over $b_0$, consistent with the findings of  M+18 that peaks have higher information content than troughs.

\begin{table}
\centering
\caption{Chosen evaluation points for the Betti functions}
\begin{tabular}{c|l}
Betti Function & Evaluation points \\
\hline\hline\\
$\regbettiKV_0$ & $-2.0 , -1.8 , -1.6 , -1.4 , -0.7 , -0.4 , -0.2$\\[1em]
$\regbettiKV_1$ & $0.1 , 0.2 , 0.4 , 0.6 , 0.8 , 1.4 , 1.6 , 1.8 , 2.3 , 2.6 ,$\\
& $ 3.0 , 3.4 , 4.0$\\[1em]
$\persbettiKV_0$ & $( -2.0 , -1.2 ),( -2.0 , -0.6 ),( -1.9 , -1.8 ),$\\
&$( -1.7 , -1.7 ),( -1.3 , -1.3 ),( -0.5 , 0.0 )$ \\[1em]
$\persbettiKV_1$ & $( -0.1 , 1.1 ),( 0.1 , 2.8 ),( 0.1 , 3.0 ),( 0.1 , 3.1 ),$\\
&$( 0.1 , 3.8 ),( 0.1 , 3.9 ),( 0.2 , 0.2 ),( 0.2 , 0.5 ),$\\
&$( 0.7 , 2.5 ),( 0.7 , 2.6 ),( 0.7 , 2.7 ),( 0.7 , 4.0 ),$\\
&$( 0.8 , 3.9 ),( 0.9 , 2.4 ),( 1.0 , 2.2 ),( 1.1 , 2.1 ),$\\
&$( 1.2 , 1.6 ),( 1.4 , 1.8 ),( 1.5 , 1.7 ),( 1.5 , 1.9 ),$\\
&$( 1.5 , 2.0 ),( 1.7 , 2.6 ),( 2.5 , 3.0 ),( 2.6 , 4.0 ),$\\
&$( 2.7 , 2.8 ),( 3.5 , 4.0 ),( 3.8 , 4.0 )$
\end{tabular}
\label{tab:evaluation_points}
\end{table}
To visualise this, we plot all features of the persistence diagram as a scatter plot, where the $x$-coordinate of a point represents the filtration value at which the respective feature is born and the $y$-coordinate represents the filtration value of its death.
\begin{figure}
  \centering
    \includegraphics[width=\linewidth]{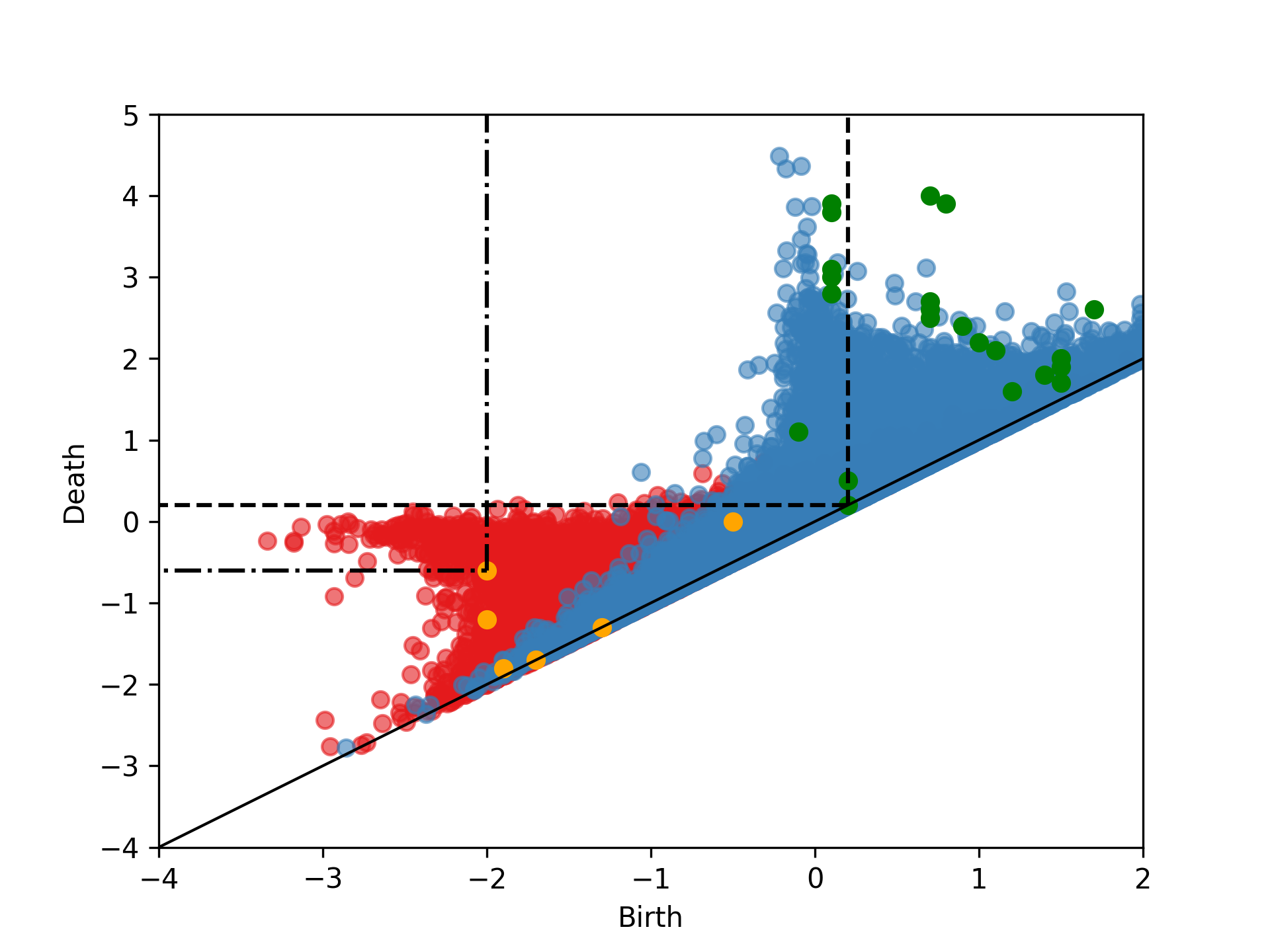}  \caption{A persistence diagram extracted from one tile $T^i$ of the SLICS, in the KV450-like setup. The scattered points represent the features in $\dgm(H_0(\mbX))$ (red) and $\dgm(H_1(\mbX))$ (blue). For each feature, the $x$-value corresponds to the $S/N$-level at which that feature is born, whereas the $y$-value corresponds to the $S/N$-level at which it disappears again. The orange and green dots mark our evaluation points of the respective persistent Betti functions $\persbettiKV_{0,1}$.  The two  dashed rectangles visualise regions in the diagram inside of which all features are counted, when computing the Betti functions at the evaluation points $(-2, -0.6)$ and $(0.2, 0.2)$.}
  \label{fig:persistence_diagram}
\end{figure}
In general, a persistent Betti function evaluated at $(x,y)$ counts all features that lie towards the upper-left of the evaluation point (see Fig.~\ref{fig:persistence_diagram}). Here, features that lie close to the diagonal arise more likely due to noise fluctuations, whereas features far away from the diagonal are statistically more significant. The final chosen evaluation points for the persistent Betti functions can be seen in Tab.~\ref{tab:evaluation_points}. Over the course of the analysis, we tested several methods of choosing evaluation points, including just setting them by hand. While different sets of evaluation points sometimes slightly change the size of the contours, they are all internally consistent and yield very similar results. In particular, completely disregarding $\regbettiKV_0$ and $\persbettiKV_0$  does not change the marginalised error on $S_8$, but helps in constraining the most remote regions of the $[\Omega_{\rm m}-\sigma_8]$  contours.

\section{Evaluating the emulator accuracy}
\label{sec:emulator_accuracy}
\begin{figure}
  \centering
  \includegraphics[width=\linewidth]{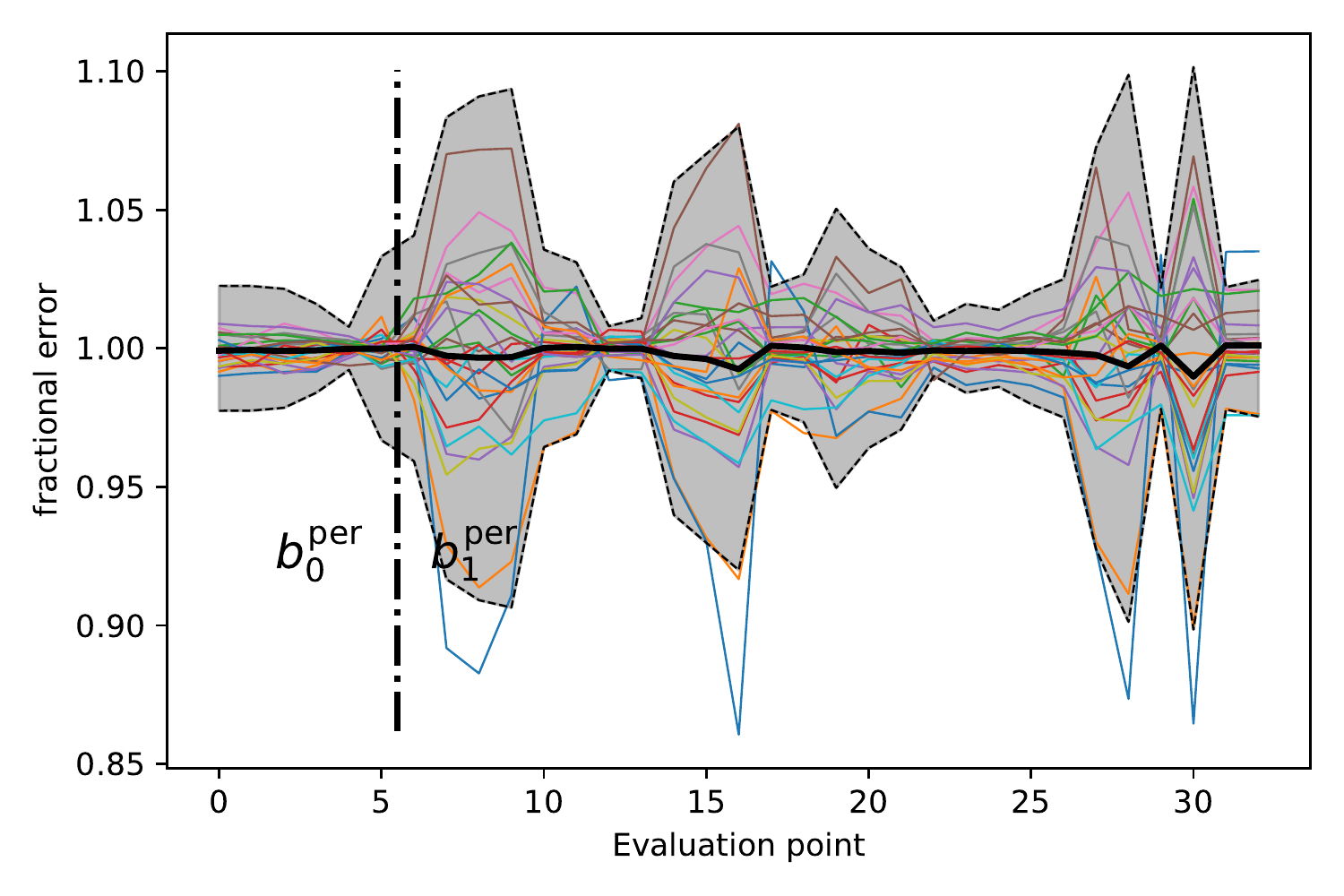}\\
  \includegraphics[width=\linewidth]{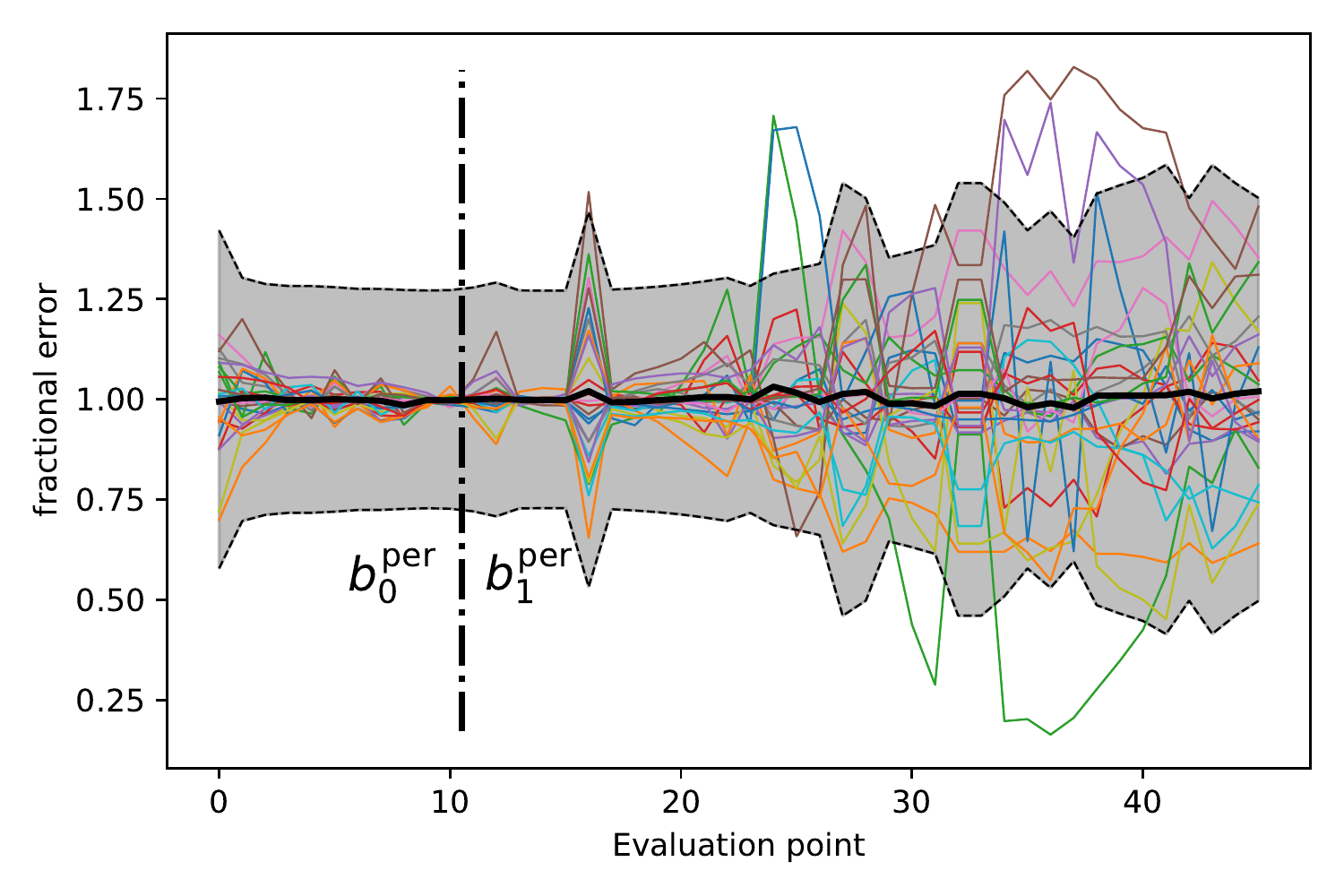}
  \caption{Relative accuracy of the emulator for predicting the persistent Betti functions of a KV450-like (top) and a \textit{Euclid}-like (bottom) survey. The thin-coloured lines represent the ratio between predicted and measured values for the 26 different cosmologies, estimated from a cross-validation test. The $x$-axis lists the points at which the functions are evaluated (see Table \ref{tab:evaluation_points} for their numerical values). The thick black line corresponds to the mean over all lines. The grey shaded area corresponds to the $1\sigma$ standard deviation of the covariance matrix extracted from SLICS.}
  \label{fig:cross_validation_results}
\end{figure}

\begin{figure*}
  \centering
  \includegraphics[width=\linewidth]{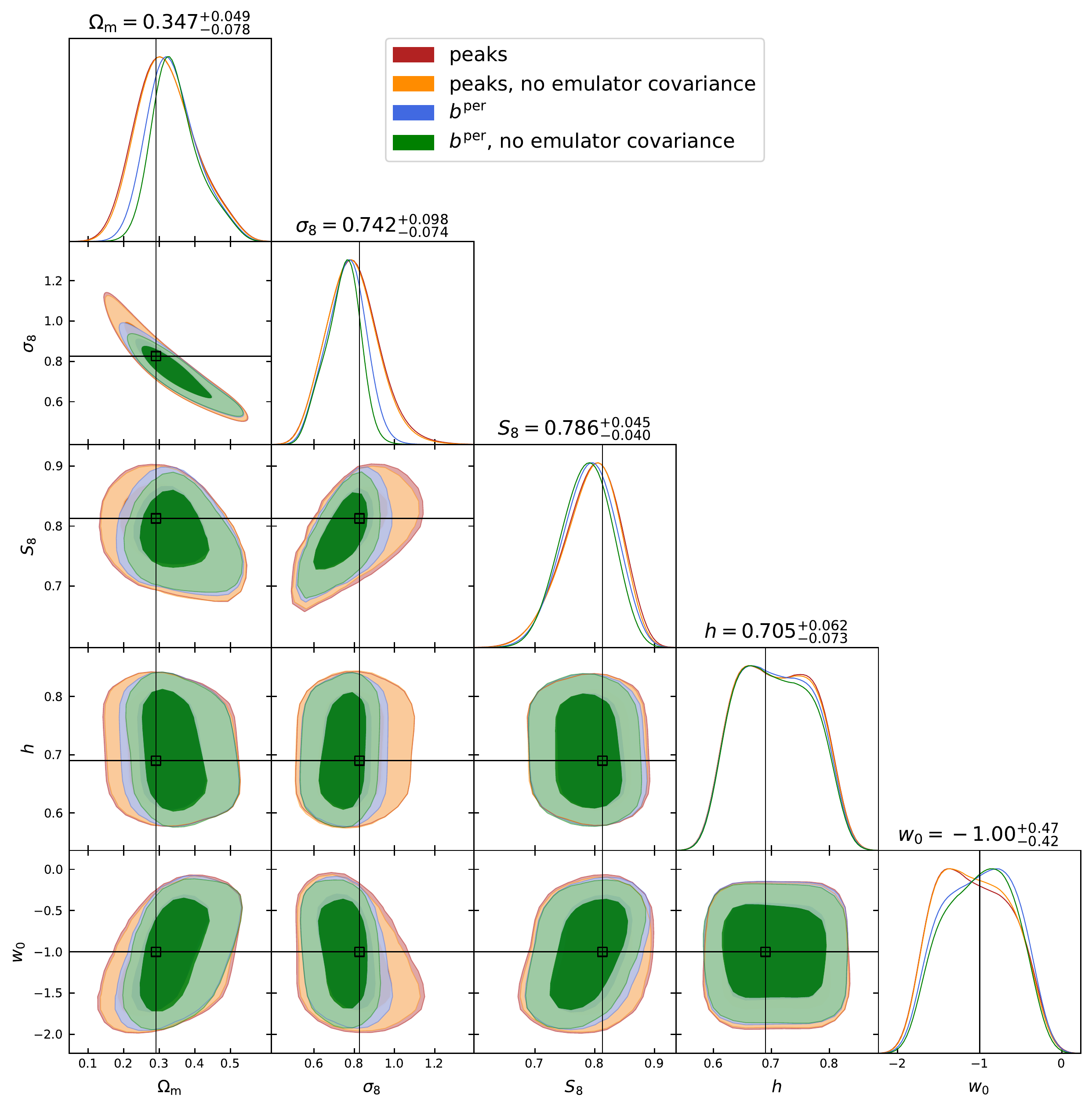}
  \caption{Parameter constraints from the persistent Betti functions and the peak statistics as presented in Fig.~\ref{fig:mcmc_result}  (blue and red, respectively), here compared to the case where the covariance of the emulator $C_\mathrm{e}$ is set to zero in the MCMC analyses (green and orange). We note that for the peaks, the emulator covariance plays a negligible role. }
  \label{fig:mcmc_no_emulator_covariance}
\end{figure*}

In this section, we further discuss the accuracy of the GPR emulator  introduced in Sect.~\ref{sec:gpr}. As described therein, we evaluate the performance via a ``leave-one-out'' cross-validation test, and present our results in Fig.~\ref{fig:cross_validation_results}  for the persistent Betti functions. 
As can be seen for the KV450-like case (upper panel), the emulator manages to predict most function values with a 5\% accuracy. While some points are less precise, with inaccuracies ranging up to 15\%, the mean inaccuracy (represented by the black line) stays well below 1\%, meaning that the emulator neither systematically under- nor over-predicts any point of the Betti functions. 
We next assess the impact of the emulator uncertainty on the inferred cosmological parameters with
an MCMC analysis in which we set the emulator covariance $C_\mathrm{e}$ to zero. As reported in Fig.~\ref{fig:mcmc_no_emulator_covariance}, this test confirms that the emulator covariance plays a non-negligible role at the moment, since in that case the constraining power is increased for all parameters, and notably the constraints on $S_8$ become tighter by $\sim\! 3 \%$.

\begin{figure*}
  \centering
  \includegraphics[width=\linewidth]{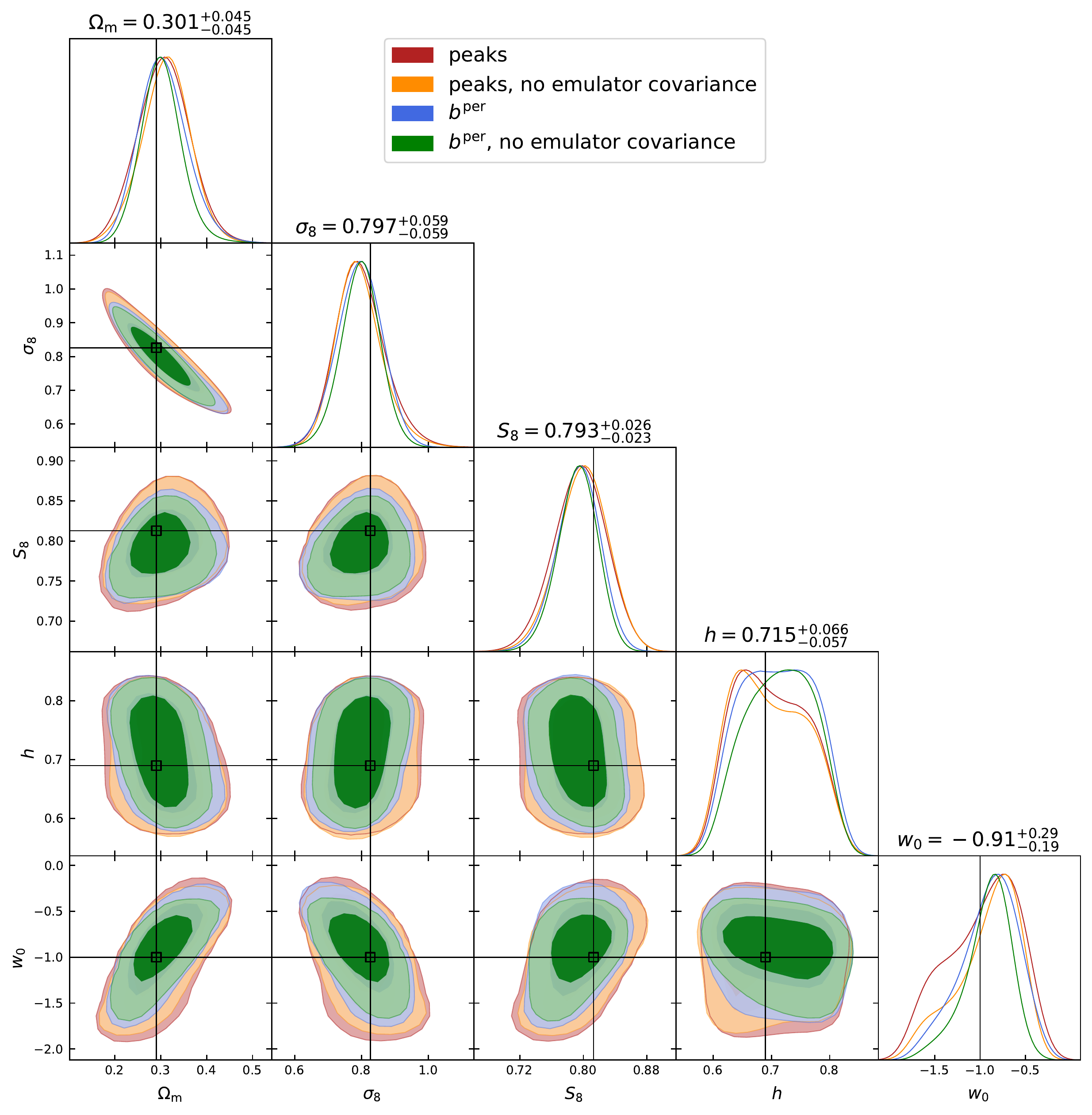}
  \caption{Same as Fig.~\ref{fig:mcmc_no_emulator_covariance}, here in the {\it Euclid}-like setup. We note that for the peaks, the emulator covariance plays a sub-dominant role. The results also indicate that, given a better emulator, persistent Betti numbers will be able to constrain the equation of state of dark energy without the need for tomographic analyses.}
  \label{fig:mcmc_no_emulator_covariance_euclid}
\end{figure*}

We applied the same procedure to the \textit{Euclid}-like simulations presented in Sect.~\ref{sec:euclid} and show the cross-validation test in the bottom panel of Fig.~\ref{fig:cross_validation_results}. Here, the emulator inaccuracies are significantly higher: while the mean fractional error in the cross-validation test still stays within a few percent, individual deviations often surpass 15\%, some points even exceeding $50\%$ fractional error. When compared with the fractional statistical error (the grey band in the figure), we observe that it is also larger compared to the KV450-like analysis. The resulting effect on the cosmological parameter analysis can be seen in Fig.~\ref{fig:mcmc_no_emulator_covariance_euclid}. While the peak statistics are only slightly affected in that case, the difference is significant for persistent Betti functions. In particular, the constraints on $S_8$, $\Omega_{\rm m}$ and $w_0$ improve by 9, 13 and 25\%, respectively.

The fact that a less noisy measurement yields a larger fractional statistical error might be counter-intuitive, but it can be understood in the following way:
In contrast to other measurements including e.g. two-point correlation functions, the presence of shape noise does not only lead to larger measurement errors, but it also changes the expectation value of the measurement itself. Due to the presence of noise, an additional number of features (or peaks) is added to each measurement. This number of pure noise peaks is, to leading order, independent of cosmology, and
since adding a constant number to a series of measurements does not affect its variance, the total sampling variance $\sigma$ is independent of the amount of noise features.
When the shape noise is lowered in our Stage IV survey simulations, the variance stays approximately constant, but the amplitude of the data vector decreases, causing the fractional statistical error to increase, as seen by comparing the size of the grey bands in both panels of Fig.~\ref{fig:cross_validation_results}. We carried out an extra test in which the intrinsic shape noise of every galaxy in the KV450-like setup was reduced by 90\%, and saw the same trend: with lower noise levels, the fractional statistical error increased almost by a factor of two in that case.

This also explains why the emulator has more difficulties in maintaining a high level of accuracy: the constant noise contribution dilutes the relative variations due to cosmology, which makes the interpolation easier for the GPR. In the {\it Euclid}-like mocks, the emulator therefore accounts for a large fraction of the error budget.  These emulator inaccuracies, and how to improve upon them, will be subject to further investigation.

While we are not yet able to reduce modelling uncertainties with the cosmo-SLICS, this will certainly be possible in the future. Fig.~A.2 of H+19 shows that for two-point correlation functions, the uncertainty of the emulator falls to the order of a percent when training it on a set of functions that has been modelled for 250 separate cosmologies. With larger projects like BACCO \citep{2020arXiv200406245A}, where  simulations are conducted for 800 different cosmologies, this accuracy is definitely achievable if a ray-tracing is performed. In that case, it is interesting to investigate the impact of a better emulator on our analysis. This can be achieved simply by performing again a cosmological parameter analysis where we set $C_\mathrm{e}=0$, as shown in Fig.~\ref{fig:mcmc_no_emulator_covariance} and Fig.~\ref{fig:mcmc_no_emulator_covariance_euclid}.

\section{Treatment of biases by noise terms}
\label{sec:systematics}
\begin{figure*}
\centering
\includegraphics[width=0.45\linewidth]{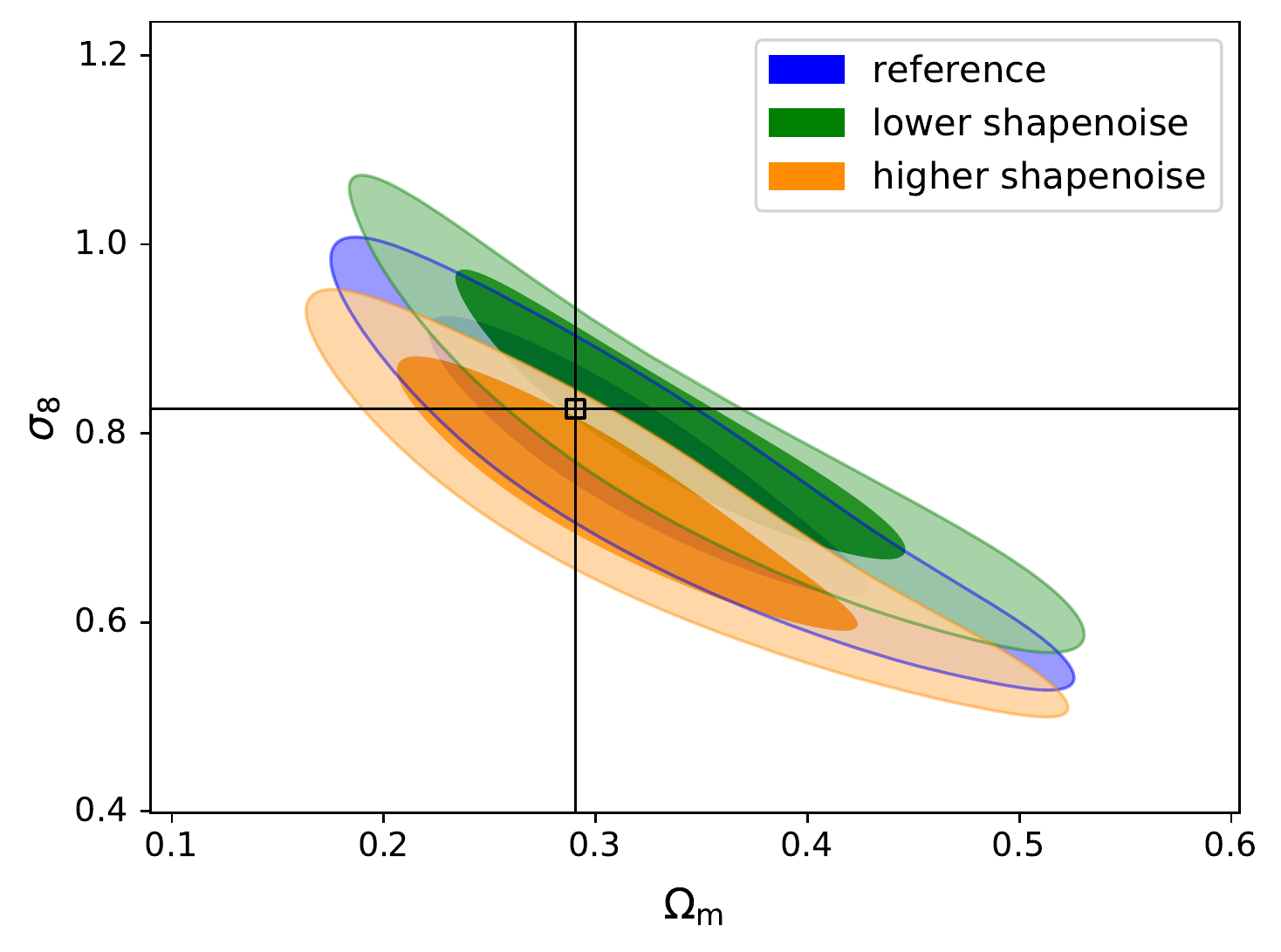}
\includegraphics[width=0.45\linewidth]{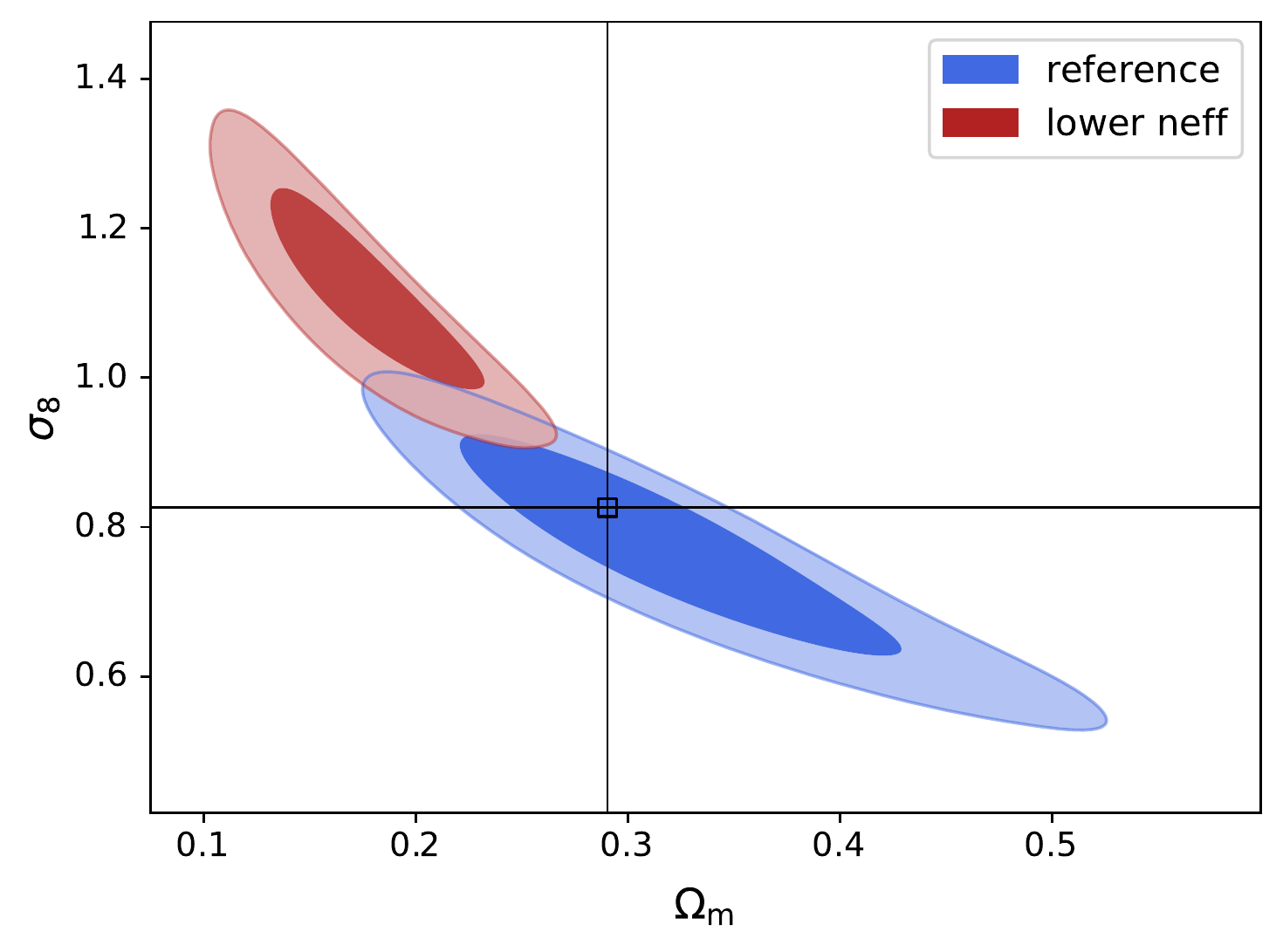}
\caption{Effect on the inferred cosmological parameter in a KV450-like survey from varying shape noise by $\pm 10\%$ (left) and lowering the number density by $10\%$ (right).  The solid lines correspond to the fiducial values of SLICS.}
\label{fig:systematics}
\end{figure*}

To test the sensitivity to number density and shape noise, we created different galaxy catalogues of SLICS, where we both increased and reduced the shape noise of galaxies by 10\%, and one where we reduced the number density by 10\% by removing galaxies at random. We used these modified catalogues to extract the mean of the persistent Betti functions $\persbettiKV$ as well as their covariance matrix. We then perform a cosmological parameter inference as in Sect.~\ref{sec:results}, where in each case the $\vec{x}_\mathrm{predicted}$ are provided by the emulator that has been trained on the unaltered cosmo-SLICS. As can be seen in Fig.~\ref{fig:systematics}, 
a 10\% change in these parameters roughly corresponds to a $\sim\!1\sigma$ shift in the posterior distribution of $S_8$. Additionally, the posterior distributions for both $\Omega_\mathrm{m}$ and $\sigma_8$ are shifted by $\sim\!3\sigma$ for the case of the lowered number density. It is therefore critical to reproduce these properties in the simulations in order to avoid very significant biases in the cosmological inference. 

In an actual cosmological data analysis, one should use mock data with a similar design to those used in this analysis, namely where galaxies have been distributed on the footprint exactly according to their positions in the data such as to reproduce number density, and additionally keeping their associate ellipticity. 
\end{appendix}

\end{document}